\documentclass[fleqn,usenatbib]{mnras}

\usepackage[T2A]{fontenc}
\usepackage{graphicx}
\usepackage{aecompl}
\usepackage[dvipsnames]{color}
\usepackage[normalem]{ulem}
\usepackage{threeparttable}
\usepackage{multirow}
\usepackage{array}

\DeclareRobustCommand{\VAN}[3]{#2}
\let\VANthebibliography\thebibliography
\def\thebibliography{\DeclareRobustCommand{\VAN}[3]{##3}\VANthebibliography}

\usepackage{graphicx}	
\usepackage{amsmath}	
\usepackage{amssymb}	

\newcommand{\degree}{^\circ}

\newcommand{\SII}{[S~{\sc ii}]}
\newcommand{\OIII}{[O~{\sc iii}]}

\newcommand{\HII}{H~{\sc ii}\ }

\newcommand{\HI}{H~{\sc i}\ }
\newcommand{\Ha}{H$\alpha$\ }
\newcommand{\Hb}{H$\beta$\ }
\newcommand{\kms}{\,\mbox{km}\,\mbox{s}^{-1}}

\newcommand{\HST}{\textit{HST}\ }
\newcommand{\SIIHa}{[S~{\sc ii}]/H$\alpha$}
\newcommand{\NIIHa}{[N~{\sc ii}]/H$\alpha$}
\newcommand{\OIIIHb}{[O~{\sc iii}]/H$\beta$}
\newcommand{\be}{\begin{equation}}
\newcommand{\ee}{\end{equation}}

\definecolor{violet}{rgb}{0.8,0,1}

\def\revone{}

\title[Ionized gas kinematics in Sextans A]{Stellar feedback impact on the ionized gas kinematics in the dwarf galaxy Sextans A.}

\author[I.S. Gerasimov et al.]{
Ivan S. Gerasimov,$^{1}$\thanks{E-mail: gerasimov.is18@physics.msu.ru}
Oleg V. Egorov,$^{2,1}$\thanks{E-mail: oleg.egorov@uni-heidelberg.de}
Tatiana A. Lozinskaya,$^{1}$
Alexei V. Moiseev,$^{3,1}$ \newauthor
Dmitry V. Oparin$^{3}$
\\
$^{1}$ Lomonosov Moscow State University, Sternberg Astronomical Institute,
	Universitetsky pr. 13, Moscow 119234, Russia
	\\
$^{2}$ Astronomisches Rechen-Institut, Zentrum f\"{u}r Astronomie der Universit\"{a}t Heidelberg, M\"{o}nchhofstra\ss e 12-14, 69120 Heidelberg, Germany
	\\
$^{3}$ Special Astrophysical Observatory, Russian Academy of Sciences, Nizhnii Arkhyz 369167, Russia
}

\date{Accepted XXX. Received YYY; in original form ZZZ}

\pubyear{2022}

\begin{document}
\label{firstpage}
\pagerange{\pageref{firstpage}--\pageref{lastpage}}
\maketitle

\begin{abstract}

Feedback from massive stars shapes the ISM and affects the evolution of galaxies, but its mechanisms acting at the small scales ($\sim 10$~pc) are still not well constrained observationally, especially in the low-metallicity environments. We present the analysis of the ionized gas (focusing on its kinematics, which were never studied before), and its connection to the massive stars in the nearby ($D \sim 1.4$~Mpc) star-forming very metal-poor ($Z\sim 0.07 Z_\odot$) galaxy Sextans~A. The analysis is based on the observations with a scanning Fabry-Perot interferometer, long-slit spectroscopy and imaging in emission lines with narrow-band tunable filters. We found 10 expanding superbubbles of ionized gas with ages of 1--3~Myr. We argue that 3 of them are probable supernovae remnants, while the pre-supernovae feedback is an important source of energy for blowing-out the remaining superbubbles. The two brightest sites of star formation exhibit signs of outflowing ionized gas, which is traced by its ionized and atomic gas kinematics and (in one case) by its emission line flux ratios. Overall, the ionized gas kinematics in Sextans~A is highly affected by the feedback from several generations of massive stars and inconsistent with the mere solid-body rotation observed in atomic hydrogen.

\end{abstract}




\section{Introduction}

Massive stars play a crucial role in the evolution of the interstellar medium (ISM) and the galaxies as a whole. Through the ionizing radiation, stellar winds and supernovae explosions are the main forces of stellar feedback \citep[e.g.][]{Krumholz2014}, \revone{which are able to} carve out low-density bubbles and shells in the ISM. The cumulative contribution of energy and momentum input from multiple stars in OB associations and young star clusters lead to the formation of large superbubbles having sizes up to several hundred pc \citep[e.g.][]{TT1988, Warren2011}. The evolution of superbubbles can lead to the redistribution of star formation activity and chemical abundance in a galaxy \cite[e.g.][]{Keller2014}. The role of stellar feedback is especially prominent in dwarf irregular (dIrr) galaxies, where the larger structures can emerge due to their thick gas-rich discs and a lack of spiral density waves. Nearby dIrr galaxies are often considered as ideal laboratories for studying the effects of massive stellar feedback on the surrounding ISM. 

The resolved observational studies of the ISM and star formation regulated by stellar feedback were performed for several well-known nearby galaxies \cite[see, e.g., ][]{Oey1996, Egorov2014, Egorov2017, Egorov2018, DellaBruna2020, Lopez2014, McLeod2019}. However, the number of such analyses for very low-metallicity galaxies ($Z<0.1Z_\odot$) is still very limited \cite[see][as some examples]{Egorov2021, Pustilnik2017, Kehrig2018}. At the same time, theoretical models and observations of stars suggest that the efficiency of the different feedback agents should vary with the metallicity. For example, both the power of stellar winds \citep{Vink2001} and the rate of SNe type II \citep{Anderson2016} decline in the low-metallicity regime. \cite{Ramachandran2019} demonstrated that the relative contribution of several feedback forces differ in superbubbles observed in \revone{the Small Magellanic Cloud (SMC)} from that in \revone{the Large Magellanic Cloud (LMC)}, and they are also different from that in the Milky Way \citep{Barnes2020}. \revone{\cite{Ramachandran2019} conclude that in the LMC the contributions from massive stars and SNe are almost equal while in the SMC SNe explosions are the dominant sources of the mechanical feedback.} It is important to check how the stellar feedback \revone{impacts the ISM} at the small scales in very low-metallicity environments.

\begin{table}
    \caption{General parameters of the galaxy Sextans A}
    \label{tab:SexA}
    \begin{tabular}{lr} \hline
        Parameters & Value  \\ \hline
        Names$^a$ & Sextans A, DDO 75, UGCA 205 \\
        Distance$^b$ & $1.38$~Mpc\\
        $M_{\rm{B}}^a$ & $-14.13^m$\\
        Linear scale & 6.7~pc~arcsec$^{-1}$ \\
        Optical radius$^c$, $R_{25}$ & 157~arcsec = 1.1~kpc \\ 
        $\log(SFR_{FUV})^d$ & $-2.04$\\
        12 + log(O/H)$^e$ & $7.54 \pm 0.06$\\ 
        $M_{\rm{HI}}^f$ & $7.24 \times 10^7 M_\odot$\\
        $M_{\rm{*}}^g$ & $13.83\times 10^7 M_\odot$\\
        Scale height$^h$ & $438\pm84$~pc\\
        Kinematic parameters$^i$:\\
        RA (J2000.0) & $10:11:01.3$\\
        DEC (J2000.0) & $-04:42:48.0$\\
        $V^{\rm{sys}}_{\rm{radio}}$ & $324 \pm 0.6$ km~s$^{-1}$\\
        Inclination, $i$ & $34^\circ$\\
        $PA_{\rm{kin}}$ & $86^\circ$\\
        $V_{\rm{rot}}$ at 50 arcsec & $13.2$ km~s$^{-1}$\\
        Vel. dispersion, $\sigma(\mathrm{H}\alpha)_{\rm{m}}^j$ & $13.6 \pm 4.6$ km~s$^{-1}$\\ 
        
    \hline
    \end{tabular}
    
    \begin{tablenotes}
    \scriptsize
    \item $^a$ LV galaxy database \\ \citep[][\url{https://www.sao.ru/lv/lvgdb/}]{Karachentsev2004}
    \item $^b$ \cite{Dalcanton2009}
    \item $^c$ HyperLEDA Database\\ \citep[][\url{http://http://leda.univ-lyon1.fr/}]{Makarov2014}
    \item $^d$ \cite{Hunter2010}
    \item $^e$ \cite{Kniazev2005}
    \item $^f$\cite{Hunter2012}
    \item $^g$ \cite{Weisz2011}
    \item $^h$ \cite{Stilp2013}
    \item $^i$ \cite{Namumba2018}
    \item $^j$ This work
\end{tablenotes}

\end{table}

This paper is dedicated to the analysis of star formation and stellar feedback in the  nearby  very low-metallicity ($Z\sim0.07Z_{\rm{\odot}}$ \citealt{Kniazev2005, Magrini2005}) galaxy Sextans~A. Its main properties are summarized in Table~\ref{tab:SexA}. This low surface brightness galaxy
belongs to the NGC~3109  association of dwarf galaxies \citep{Tully2006} and is weakly interacting with two nearby galaxies of comparable or higher masses (Sextans~B and NGC~3109, \revone{separated from Sextans~A by 280~kpc and 500~kpc, respectively \citealt{Bergh1999}.}) However, the whole association is relatively isolated as it is located near the periphery of the Local Group. The most metal-poor massive stars known in the Local Group galaxies have been observed in Sextans~A \citep{Hosek2014, Garcia2019, Kaufer2004}. 
 
According to the analysis of the star formation history based on the \revone{Hubble Space Telescope (\textit{HST})} data, Sextans~A has been intensively forming stars during the last $\sim60$~Myr, while more quiet star formation activity has been taken place during the last $1-2.5$~Gyr \citep{Dolphin2003, Dohm2002}. The star formation activity was possibly triggered by the interaction with the other members of the NGC~3109 group, or with the Milky Way, that indeed should happen several Gyr ago and led to the low surface brightness stellar tidal tails at the periphery of the galaxy's disc \citep{Bellazzini2014}. Meanwhile, the solid-body rotation of \HI disc and the absence of the prominent radial motions there allow to suggest that the current star formation is mostly regulated by internal processes \citep{Ott2012, Namumba2018}. 

\begin{figure}
    \centering
    \includegraphics[width=\linewidth]{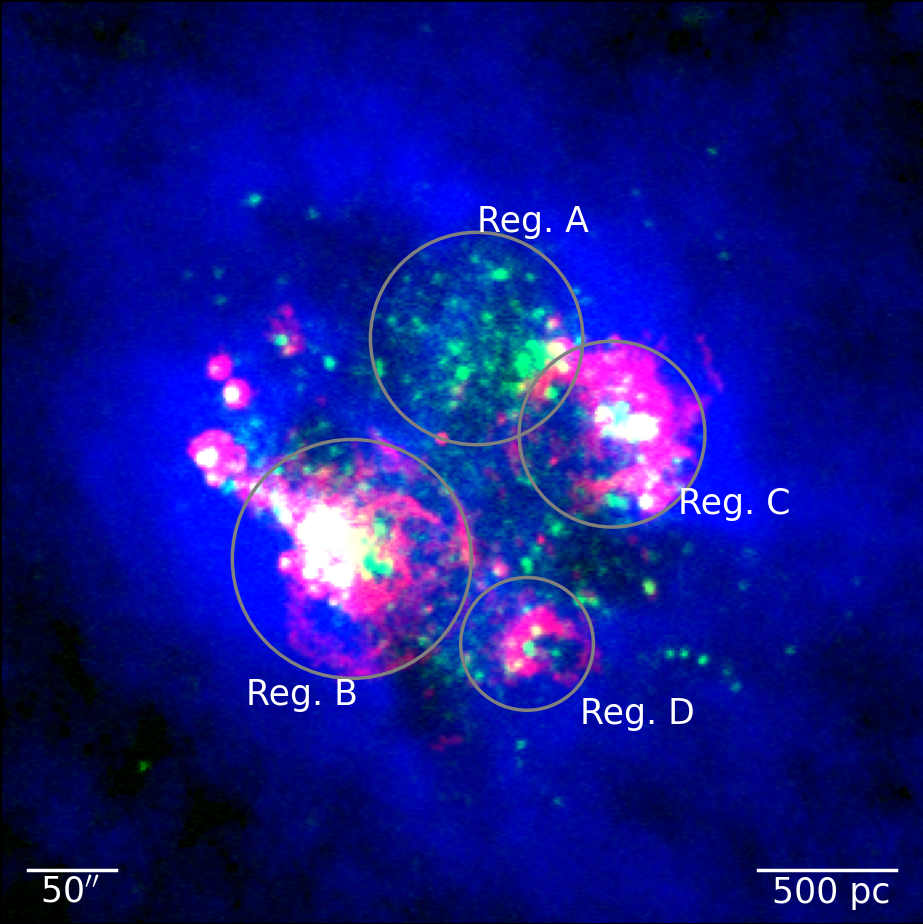}
    \caption{False-color image of Sextans~A: red,
green and blue channels correspond to emission in H$\alpha$ (KPNO), FUV (GALEX) and
\HI 21 cm (VLA), respectively.
Grey ellipses show the location of four main star-forming complexes in the galaxy.
}
    \label{fig:map}
\end{figure}

The ongoing star formation in Sextans~A takes place in 4 complexes clearly distinguishable in far-ultraviolet (FUV; see Fig.~\ref{fig:map}). The oldest one (Region A in Fig.~\ref{fig:map}) has an age of $\sim 400$~Myr \citep{Dohm2002} and does not reveal any associated \Ha emission. The brightest complex of star formation in \Ha (Region B) lies south-east and has an age of $\sim 200$~Myr (according to \citealt{Camacho2016}) and still is more active than the second-brightest north-western Region C, which is much younger with the age of about 20~Myr. The southern small complex (Region D) is not very actively forming stars -- \citet{Garcia2019} identified 4 OB stars there and there is no related OB association or star cluster, also optical images in \Ha reveal much fainter surrounding \Ha emission than in complexes B and C. \citet{Garcia2019} suggested that star formation there might process directly from \HI like in XUV discs. Very low efficiency of star formation was reported by \cite{Shi2014}.

One intriguing feature that attracted our particular attention to this galaxy is its appearance in the \HI 21~cm line. The atomic hydrogen is distributed around the stellar body and star-forming regions in a form of a single supershell with a radius of $R\sim850$~pc (see Fig.~\ref{fig:map}). There are only few galaxies with such \HI morphology known within nearby  $\sim 10$~Mpc  (including Holmberg~I --- \citealt{Egorov2018}, DDO~88  ---  \citealt{Simpson2005}, M81 dwarf~A  --- \citealt{Sargent1983}, Sag~DIG  --- \citealt{Young1997}), and Sextans~A is the second nearest of them (after Sag~DIG). \revone{In contrast with the galaxy Holmberg~I, which appearance is similar to Sextans~A}, the ongoing star formation activity is not embedded in the walls of the \HI supershell, but rather located towards its inner edge. The morphology of the ionized gas represents several extended filaments and shell-like structures related to three star-forming complexes in the galaxy. \cite{vanDyk1998} proposed that star formation activity began $\sim 50$~Myr ago in the centre of the galaxy and then propagated towards its periphery producing the supershell. However \citet{Dohm2002} later revealed in the rims of \HI supershell older stars than in its centre that contradicts this scenario. Nevertheless, the energy input of multiple 
generations of massive stars is sufficient to blow-out the central \HI supershell, especially considering that it might consist of several overlapping smaller structures \citep{Warren2011}.

Because of the typically better spatial resolution (and also a wider range of possible diagnostics), the impact of the stellar feedback onto the ISM at the  scales of about 10 pc is usually better traced by the observations of the ionized gas. However, the kinematics and excitation conditions of the ionized gas in Sextans~A, as well as its relation with the massive stars and atomic gas, remain almost unexplored. The only characterization of the ionized gas kinematics was published by \cite{Hunter1997} who obtained 5 long-slit echelle spectra crossing the prominent \Ha filaments and shells. They found no signs of expansion for the \Ha shells surrounding the complexes B, C and D (Fig.~\ref{fig:map}), nor a difference in velocity of the filaments and the \HII regions in their data, but identified the expanding bubble associated with the brightest \HII region in the complex B.  

The information on the 2D distribution of the ionized gas kinematics is still required for the understanding of the role of stellar feedback shaping the ISM of Sextans~A at small scales. To fill this gap in our knowledge of this nearby and interesting object, we observed it with a scanning Fabry-Perot interferometer (FPI) at the 6-m Big Telescope Alth-azimuthal (BTA) of the Special Astrophysical Observatory of the Russian Academy of Sciences (SAO RAS). These data are accompanied also by the BTA long-slit spectrum and images with tunable narrow-band filter obtained at the 2.5-m of the Caucasian Mountain Observatory   of the Sternberg Astronomical Institute of the Moscow State  University (SAI MSU). In this paper, we analyse the properties of the ionized gas and the interplay between massive stars and the ISM in Sextans~A.

The paper is organized as follows. In Section~\ref{sec:obs} we describe the new observations and archival data used for the analysis. Section~\ref{sec:analysis} presents the analysis of the obtained data, namely the analysis of the ionized and atomic gas velocity fields (Sec.~\ref{sec:velocities}), the identification of the expanding ionized superbubbles (Sec.~\ref{sec:superbubbles}), the ionization state of the ISM in the star-forming regions (Sec.~\ref{sec:ion_state}). In Section~\ref{sec:discussion} we evaluate the obtained results to make conclusions on the energy balance between the ionized gas and stars (Sec.~\ref{sec_an_energy}), possible local imprints of a stellar wind from very young massive star in the complex~D (Sec.~\ref{sec:disc_star}) and on the presence of stellar feedback-driven outflows in the Sextans~A (Sec.~\ref{sec:disc_outflow}). Section~\ref{sec:summary} summarizes our main findings.

\section{Observation and data reduction}
\label{sec:obs}

Our analysis is based on the optical data obtained at the 6-m telescope BTA of SAO RAS and the 2.5-m telescope of SAI MSU, as well as on the archival multiwavelength data. The log of the  observations is given in Table~\ref{tab:obs_data}. 

\begin{table*}
	\caption{Log of observational data}
	\label{tab:obs_data}
	\centering
	{\footnotesize
	\begin{tabular}{llrllllll}
		\hline
		Data set       & Date of obs    & $\mathrm{T_{exp}}$, s & FOV                             & $''/px$             & $\theta$, $''$   & sp. range          & $\delta\lambda$, \AA        \\ \hline
		FPI \#1  & 2017 Dec \revone{12} & $40\times120$  & {$6.1\arcmin\times6.1\arcmin$} & {0.71} & 3.0              & {8.8~\AA\, around \Ha}  &{0.48} \\
		FPI \#2  & 2017 Dec \revone{12} & $40\times150$  & {$6.1\arcmin\times6.1\arcmin$} & {0.71} & 2.6              & {8.8~\AA\, around \Ha}  &{0.48  } \\
		FPI \#3  & 2020 Feb 17 & $40\times150$  & {$6.1\arcmin\times6.1\arcmin$} & {0.71} & 1.5              & {8.8~\AA\, around \Ha}  &{0.48  } \\
		LS PA=74 & 2020 Dec \revone{24} & 4500 & {$1\arcsec\times6\arcmin$} & {0.40} & 1.5 & {3710--7180}  & $4.8$ \\
		MaNGaL & 2021 Apr 16 & 1500 & {$5.6\arcmin\times6.1\arcmin$} & {0.33} & 2.4 & \OIII$\lambda5007$   & $13$ \\
		MaNGaL & 2021 Apr 16 & 1800 & {$5.6\arcmin\times6.1\arcmin$} & {0.33} & 2.0 & \OIII{} continuum & $13$ \\
		MaNGaL& 2022 Apr 24 & 3600 & {$5.6\arcmin\times5.6\arcmin$} & {0.33} & 1.2 & \SII$\lambda6717$  & $13$ \\
		MaNGaL& 2022 Apr 24 & 3600 & {$5.6\arcmin\times5.6\arcmin$} & {0.33} & 1.2 & \SII$\lambda6731$  & $13$ \\
		MaNGaL& 2022 Apr 24 & 3600 & {$5.6\arcmin\times5.6\arcmin$} & {0.33} & 1.2 & \SII{} continuum  & $13$ \\
		\hline
	\end{tabular}}
	\begin{tablenotes}
	\scriptsize
	\item \revone{$\mathrm{T_{exp}}$ is the exposure time;}
	 \revone{FOV is the field of view;}
	 \revone{$''/px$ is pixel size on the final images;}
	\item \revone{$\theta$ is the final angular resolution;} 
	 \revone{$\delta\lambda$ is the final spectral resolution.}
	\end{tablenotes}
\end{table*}

\subsection{Scanning FPI observations}
\label{sec_obs_ifp}

The observations were obtained at the prime focus of the 6-m telescope of SAO RAS using a scanning FPI mounted inside the SCORPIO-2 multi-mode focal reducer \citep{scorpio2}.

The operating spectral range around the \Ha emission line was cut by a bandpass filter with a $\mathrm{FWHM}\approx14$~\AA\ bandwidth. We have consecutively obtained 40 interferograms at different distances between the FPI plates. 
The data reduction was performed using a software package running in the \textsc{idl} environment. For a detailed review  of the data reduction algorithms and software see \citet{Moiseev2021}.  After the initial reduction, sky line subtraction, photometric and seeing corrections made using the reference stars,
and wavelength calibration, the observational data were combined into data cubes, where each pixel in the field of view contains a 40-channel spectrum. The width of the instrumental profile corresponds to the velocity resolution, i.e., about $22\kms$.

We observed the galaxy in two overlapped fields. Complexes B and D were exposed in 2017 at two position angles to remove the parasitic ghost reflection \cite[see][for more details]{Moiseev2008}. Complex C was exposed only with one position angle. After data reduction, we constructed a mosaic of these two fields. \revone{Two data cubes were convolved with a Gaussian to reach the same angular resolution as in the data cube FPI~\#1 with the worst seeing (see Table.~\ref{tab:obs_data})}.

The result of the FPI observations and data reduction is a large--scale data cube containing 40-channel spectra in the region around the red-shifted \Ha line. The analysis of the emission line profiles is carried out using single-component Voigt fitting \citep{Moiseev2008}, which yields flux, line-of-sight velocity and velocity dispersion (corrected for instrumental broadening) as an output. The measured values of the \Ha velocity dispersion were additionally corrected for natural and thermal broadening by subtracting in quadrature the $\sigma_\mathrm{th}=9.6\ \kms$ (\revone{assuming $T_\mathrm {e} = 10\,000$~K}). 
To calibrate to the absolute intensities we use the same procedure as described in section~\ref{sec_obs_other} for archival data.


\subsection{Long-slit spectroscopic observations}
\label{sec_obs_ls}

The long slit observations were carried out with the same SCORPIO-2 multi-mode focal reducer at the   6-m telescope of SAO RAS. 

The data reduction was performed in a standard way using the SCORPIO-2 \textsc{idl}-based  pipeline as described in \citet{Egorov2018}.  The observations of the spectrophotometric standards G191 B2B at a close zenith distance immediately before the Sextans~A observations were used to calibrate its spectra to an absolute intensity scale.

To measure the fluxes of emission lines our own \textsc{idl} software  based on the \textsc{mpfit} \citep{mpfit} routine was used. Gaussian fitting was applied to measure the integrated line fluxes of
each studied region. For estimating the final uncertainties of the line fluxes we added in quadrature the errors propagated through all data reduction steps to the uncertainties returned by \textsc{mpfit}. We didn't perform any modelling or subtraction of the underlying stellar population because of its negligible contribution to the emission spectra in Sextans~A \revone{(measured equivalent width EW(H$\alpha$)$ > 100$~\AA\ in the \HII regions)}.

\subsection{Narrow-band imaging}
\label{sec_obs_figs}
Narrow-band optical images of Sextans~A in the \revone{\OIII$\lambda5007$\ and \SII$\lambda6717$ and $\lambda6731$\ emission lines}  were taken at the Nasmyth focus of the 2.5-m telescope of SAI MSU with the tunable filter imager MaNGaL \citep[Mapper of Narrow Galaxy Lines,][]{mangal2020}. In MaNGaL, the scanning FPI working at low order of interference acts as a narrow band filter with $FWHM\approx13$\AA. We set the  central wavelength of the filter according to the mean line-of-sight velocities of the studied emission lines 
and the nearest continuum at the distances 30--50\AA\ on  both sides of the lines. 
Data reduction was performed in the same way as described in our previous papers \citep[e.g.][]{Oparin2020AstBu..75..361O}. Images of the standard star AGK+81D266 taken at a close zenith distance immediately after or before the galaxy observations were used to calibrate the continuum-free emission line images to the absolute intensities. The observations in the \OIII\ emission line were obtained under non-photometric conditions, therefore a shift of a photometric zero-point is possible.

\subsection{Other observational data used}
\label{sec_obs_other}

We used the archival JVLA data in \HI 21 cm line from the LITTLE THINGS survey \citep{Hunter2012} to study the \HI gas distribution and kinematics. In this work, we analyse the natural-weighted (NA) data cube which has a velocity scale of $2.6 \kms$ per channel and angular resolution of $\theta_{NA}=11.8\times9.5$~arcsec. \revone{We used the published maps of zero and first statistical moments showing the \HI 21 cm flux and line-of-sight velocity}, respectively.

From the scanning FPI observations we extracted the map in \Ha line that is deeper than the previously obtained images available in archive. However, its angular resolution (limited by seeing) is rather low, so 
we decided to use also the archival narrow-band image in \Ha line from the 4-m telescope Kitt Peak National Observatory \citep{Masseylines, Masseyubvri} to trace the small-scale distribution of the ionized gas in the galaxy. The underlying stellar continuum was subtracted using the image in $R$ band from the same data set. We use a calibrated image from \citet{Kennicutt2008} to calibrate the KPNO data. For this purpose, we aligned both maps and then used the peak distribution of their ratio to normalize the KPNO image. 

In Fig.~\ref{fig:map} we show the distribution of the recent star formation activity in Sextans~A by using the far UV images obtained with the GALEX space telescope as a part of the LITTLE THINGS survey.

\section{Analysis}
\label{sec:analysis}

\subsection{Velocities of the ionized and atomic gas.}
\label{sec:velocities}

\begin{figure*} 
    \centering
    \includegraphics[width = \textwidth]{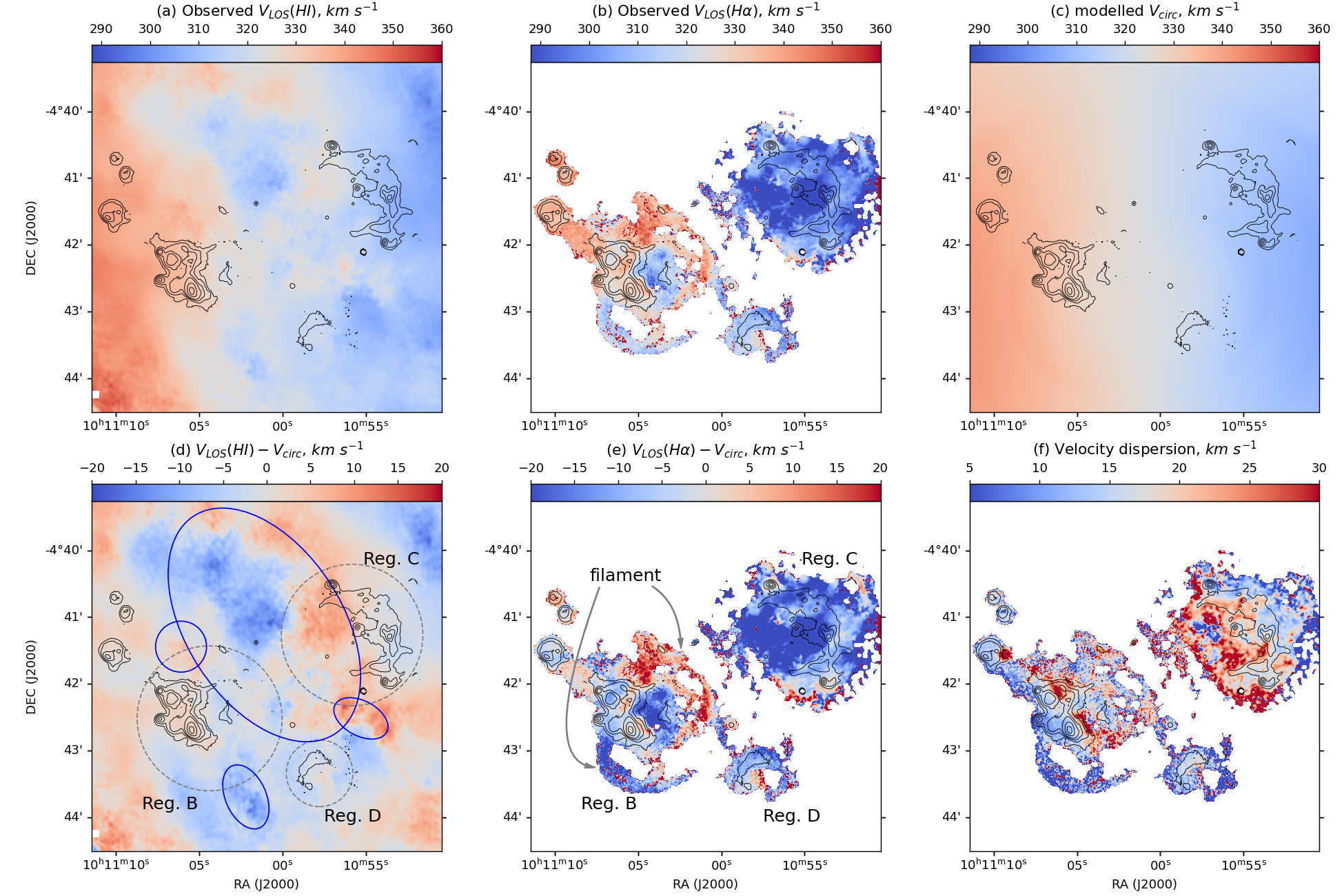}
    \caption{The velocity field in \HI 21~cm and H$\alpha$. Panels (a) and (b) -- the observed velocity in \HI and H$\alpha$, respectively. Panel (c) shows the modelled velocity field of circular rotation for Sextans~A. Panels (d) and (e) -- subtraction of the circular rotation model for \HI and H$\alpha$, respectively. Blue ellipses on panel (d) show the locations of the atomic gas shells identified by~\citep{Pokhrel2020}. We also mark the star-forming regions discussed in the paper with grey dashed ellipses. (f) shows dispersion velocity field in H$\alpha$. Contours on each panel are lines of constant \Ha surface brightness.}

    \label{fig:vel_map}
\end{figure*}

In this Section we compare the general atomic and ionized gas kinematics in the galaxy Sextans~A. Our analysis is focused on the star-forming regions located on the rim of the supershell of neutral hydrogen, but we considered the whole extent of the \HI data to obtain the circular rotation model (see below). The  first-moment map from the \HI 21~cm data is considered as the observed velocity field in this line and shown in Fig.~\ref{fig:vel_map}a. For the ionized gas, we analysed the velocity field obtained from the FPI data cube as described in Sec.~\ref{sec_obs_ifp}  (Fig.~\ref{fig:vel_map}b).

Stellar feedback from massive stars created a very complex structure and kinematics of the ISM that is mixed with the disc rotation pattern. To better understand the local gas kinematics we first excluded the component associated with the galaxy rotation following the `derotation' procedure described in~\citet{Egorov2014}. 
The model of a regular rotation of the galaxy was recovered from the \HI 21~cm data and the kinematics parameters of Sextans~A presented by \citet{Namumba2018}. They derived the kinematical positional angle, inclination, systematic velocity, and position of the rotation centre (see Table~\ref{tab:SexA}). \citet{Namumba2018} demonstrated that these values do not vary significantly inside $R < 350\arcsec$. We thus fixed these parameters and iteratively fitted the tilted-ring model to the observed \HI velocity field as described in \citet{Moiseev2014}. The free parameter of the model was the radial profile of the rotation velocity. On the second iteration, we masked all the regions with the residual velocities $> 7.5 \kms$ which might be significantly affected by the non-circular motions due to stellar feedback or other processes. The resulting model of a circular rotation is shown in Fig.~\ref{fig:vel_map}c.

\begin{figure*}
    \centering
    \includegraphics[width = \textwidth]{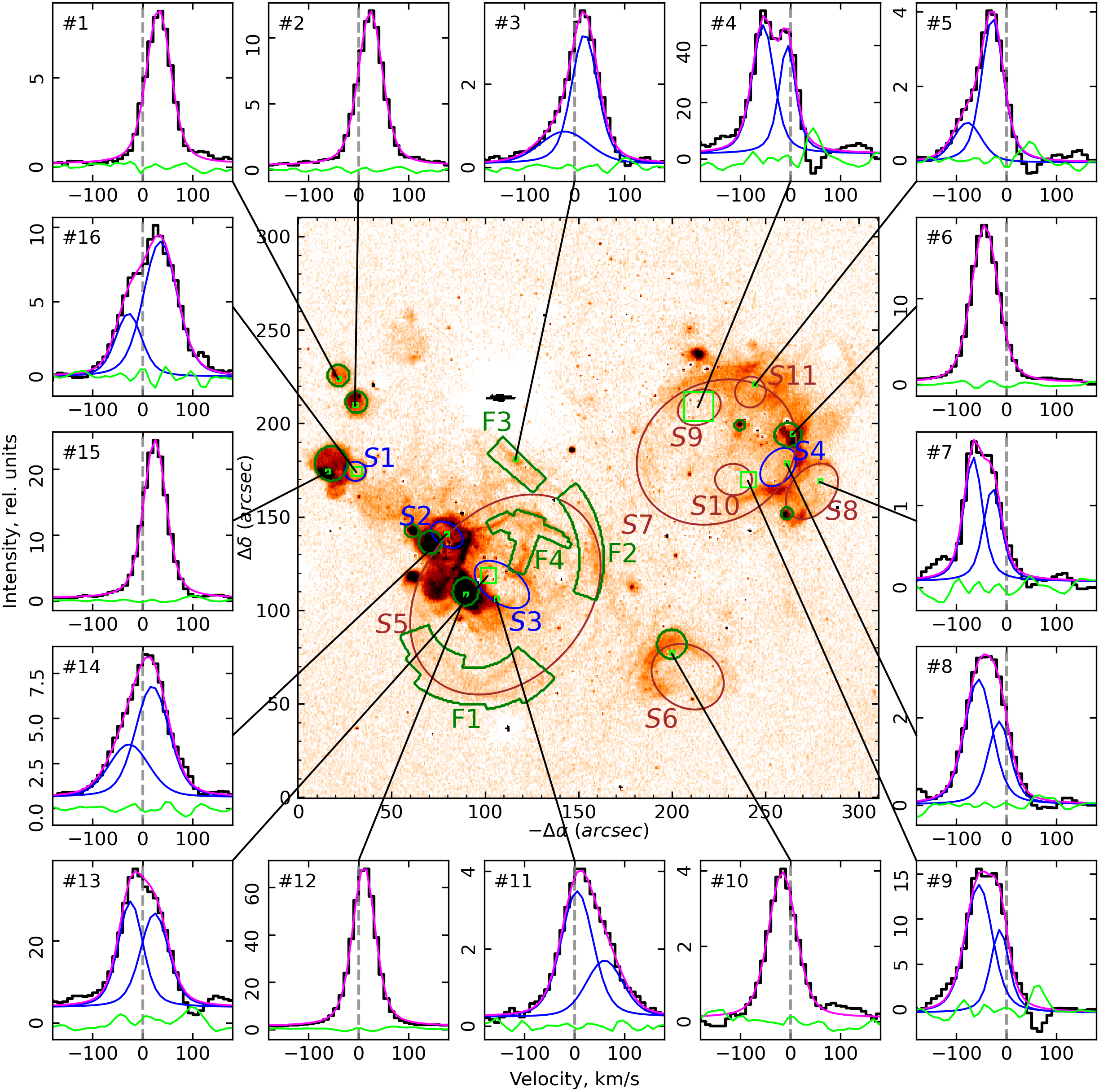}
    \caption{\Ha image of Sextans~A. \Ha line profile examples were obtained by integration inside the lime borders in the image. The result of 1 or 2-component Voigt fitting is shown: blue colour denotes the individual components, while the fitting model is shown in magenta colour. The green colour is used to show the residuals after subtraction of the model from the observed line profile. \revone{The zero-velocity line is shown by a grey dashed line.}  Blue ellipses in the image mark the positions of superbubbles identified with $I-\sigma$ diagram (Fig.~\ref{fig:i-sigma}). Brown ellipses correspond to the superbubbles identified by their morphology. \revone{Green areas and circles denote the filaments and \HII regions, respectively,} considered further in Sec.~\ref{sec:ion_state}.} 
    \label{fig:spectr_example}
\end{figure*}

Panels (d) and (e) in Fig.~\ref{fig:vel_map} show the residual line-of-sight velocities in the \HI 21~cm and \Ha lines, respectively, after subtraction of the  model of a circular rotation from the observed velocity fields in these lines. The non-circular motions in \HI line are very significant in the Sextans~A (especially in the regions A and C) and cannot be completely explained by the obtained simple model of a circular rotation, though the described procedure made the distribution of the \HI line-of-sight velocity significantly more shallow. We note that the most significant residuals are observed towards the \HI supershells identified by \cite{Pokhrel2020} and thus probably related to the stellar feedback impact. 
The qualitative picture of the \Ha line-of-sight velocity distribution is less affected by the subtraction of the rotational component. 

We emphasize several features in the resulting rotation-subtracted velocity fields in \HI and \Ha lines:

\begin{enumerate}
    \item A shallow distribution of the residual velocities in the \HI line in the brightest star-forming region (Reg. B) is accompanied by the strong variation of the line-of-sight velocities in the \Ha line (visible also in the original  velocity field). This complex B consists of several bright \HII regions surrounded by a net of faint extended filaments (see Figs.~\ref{fig:map},\ref{fig:spectr_example}). Both southern and northern filaments of the ionized gas have significant negative and positive velocities. The southern filament (F1 in Fig.~\ref{fig:spectr_example}) has also blue-shifted velocities in \HI line, which means the ionized and neutral gas can be co-spatial there. In contrast, the northern filament (F2 and F3 in Fig.~\ref{fig:spectr_example}) is clearly red-shifted not only with respect to the rotation model but also to the H~\textsc{i}. One of the possible explanations can be that the \Ha emission there originated from the background material (e.g., the rear wall of the central \HI supergiant shell) illuminated by the ionizing radiation or impacted by the shocks produced by the star clusters in the complex~B. 
    \item 
    \revone{Most of the blue velocities in the \Ha line in the complex B are observed near the brightest \HII region.} Meanwhile, the velocity of the \HI there is close to zero. This is indicative of the possible presence of the expanding ionized gas superbubble (see Sec.~\ref{sec:superbubbles}), or of the outflow in this region. 
    \item A similar picture is observed in the central part of the region~C where the blue-shifted \Ha velocities are even higher. At the same time, the atomic hydrogen is mostly red-shifted there. Such a strong disagreement between the ionized and atomic gas kinematics points to their emission originating in different places. While the \HI emission can come mostly from the rear wall of the central \HI supergiant shell, the \Ha emission might be associated with the ionized gas outflow in the low-density environment in the centre of the complex~C. 
\end{enumerate}

As follows from these, local kinematics of the ionized and neutral gas significantly differ in multiple regions of Sextans~A. At the same \revone{time} the residual velocities of \HII regions are close to zero in both \HI and \Ha lines. In contrast with previously reported findings of the ionized gas kinematics in this galaxy based on the long-slit observations \citep{Hunter1997}, our FPI data demonstrate that it is strongly perturbed, especially in the star-forming complex B. 

\begin{figure*}
    \centering
    \includegraphics[width = \textwidth]{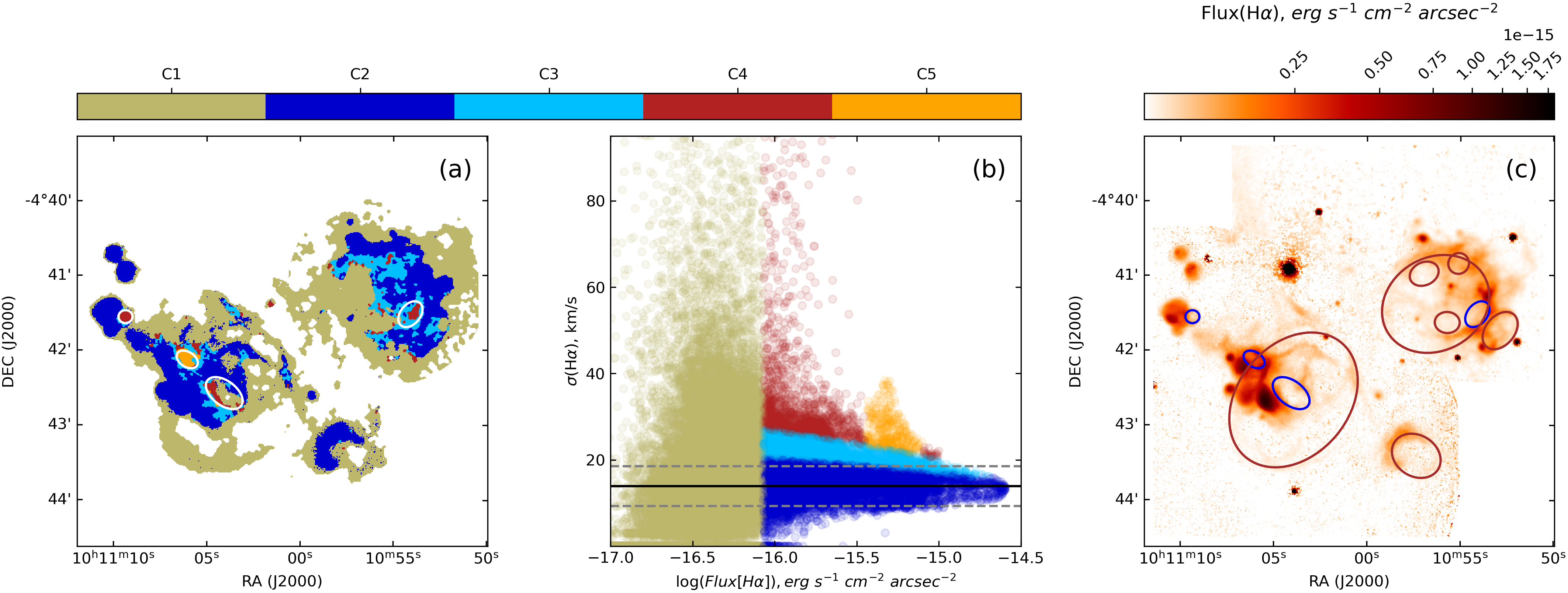}
    \caption{{\bf Panel (a)} classification map according to $I-\sigma$ diagram presented on panel (b). Different colours encode several types of regions denoted on the colour bar (colours are the same in panels (a) and (b)). White ellipses are the superbubble candidates that have elevated $\sigma(H\alpha)$. {\bf Panel (b)} -- $I-\sigma$ diagram of the \Ha line showing the pixel-by-pixel dependence of the $\sigma(H\alpha)$ on the logarithm of the line flux F(H$\alpha$). The description of each class is given in the text. {\bf Panel (c):} -- \Ha map of Sextans~A. Blue ellipses are \revone{the expanding bubble or superbubble} candidates identified with $I-\sigma$ diagram. Superbubble candidates identified by other methods were overlaid as brown ellipses.}
    
    \label{fig:i-sigma}
\end{figure*}

\subsection{Small-scale gas kinematics and expanding superbubbles}
\label{sec:superbubbles}

\subsubsection{\revone{Multi-component \Ha line profiles}}
\label{sec_halpha}
Several shell-like structures of different sizes are clearly visible in the distribution of the ionized gas in Sextans~A. From visual inspection of the \Ha image, we identified several such regions shown by brown and blue ellipses in Fig.~\ref{fig:spectr_example}, which we consider further as probable expanding superbubbles. In previous Section~\ref{sec:velocities} we found that the line-of-sight velocity towards several of them (S3 and S7) is indicative of the ionized gas outflow as these regions exhibit strong blue-shifted non-circular motions. Similarly, we can suggest that the region S6 in Fig.~\ref{fig:spectr_example} (encircling the star-forming complex~D) also shows evidence of expansion as we observe the significant red-shifted residual velocities towards its centre (see Fig.~\ref{fig:vel_map}e). In contrast with the two previous regions, here we observe similar red-shifted non-circular motions in both \Ha and \HI lines. The dominating red-shifted velocities might be indicative of a blister-like nature of the region S6 (and the entire complex~D) -- we can observe here mostly receding side of a superbubble due to the inhomogeneous ISM. 

From the analysis of the \Ha line profile from the FPI data we can judge on the presence of a secondary component related to the superbubble. The signal-to-noise towards the centre of the regions S6 and S7 is insufficient for this, and it is not possible to isolate components relating to S5 (encircled by northern and southern filaments) due to its large size and complex structure. We show the line profiles extracted from all the other superbubble candidates in Fig.~\ref{fig:spectr_example}. All these regions exhibit \revone{asymmetrical line profiles, which can be decomposed into} at least two kinematically distinct components\footnotemark{} corresponding to the possible expansion velocities of superbubbles of $\sim 20-30\ \kms$ (based on the separation between the individual Voigt components; see below). The examples of the \Ha line profile extracted from the \HII regions (shown by green circles) are also shown in Fig.~\ref{fig:spectr_example} as a reference -- they are always single-component and demonstrate lower $FWHM$. \footnotetext{\revone{We fit up to three Voigt-components into the line profiles. Accepted models are those with a minimal number of components leading to residuals within 3$\sigma$.}}


\subsubsection{\revone{$I-\sigma$ diagram}}
\label{sec_isigma}
We analysed the relative spatial distribution of the line-of-sight velocity dispersion in \Ha line (shown in Fig.~\ref{fig:vel_map}f) in order to check if the spatial extent of the identified kinematical features (broadened and two-component line profiles) correlates with the shell-like morphology of the superbubble candidates. While the high velocity dispersion is usually observed in the diffuse ionized gas (DIG)  and complicates the analysis of the velocity dispersion maps \citep{Moiseev2015}, the expanding structures appear much better on so-called $I-\sigma$ diagram (\Ha flux vs line-of-sight velocity dispersion for each individual pixel) as triangle-shape features \citep{Munoz1996, Moiseev2012}. We followed the approach presented in \cite{Egorov2021} to construct the $I-\sigma$ diagram (Fig.~\ref{fig:i-sigma}b) and corresponding classification map (Fig.~\ref{fig:i-sigma}a) for Sextans~A. First, we masked out all pixels with $S/N < 5$ in the \Ha flux distribution obtained from FPI data (Fig.~\ref{fig:i-sigma}c). All the other pixels are shown on $I-\sigma$ diagram. 
\revone{We excluded from the further analysis all the pixels with \Ha flux below the median value among all shown on $I-\sigma$ diagram -- DIG emission is probably dominating in the corresponding regions.}
This low surface brightness component is marked as C1 in Fig.~\ref{fig:i-sigma}. From remaining data points, we derived the intensity-weighted average velocity dispersion of ionized gas in the galaxy: $\sigma_\mathrm{m} \pm \sigma_\mathrm{std}= 13.6 \pm 4.6\ \kms$. This value characterizes the mean velocity dispersion of the ionized gas in Sextans~A. Blue color (C2) in Fig.~\ref{fig:i-sigma} corresponds to those pixels with normally-distributed velocity dispersion for a particular \Ha intensity ($\sigma(\mathrm{H}\alpha) < \sigma_m + 1.5 \sigma_{std}$). These pixels correspond to \HII regions or other bright emission-line structures. Regions of interest are those having high velocity dispersion for their intensity: C3 ($\sigma(\mathrm{H}\alpha) = \sigma_\mathrm{m} + (1.5 - 3)\times \sigma_\mathrm{std}$), C4-C5 ($\sigma(\mathrm{H}\alpha) > \sigma_\mathrm{m} + 3 \sigma_\mathrm{std}$). The regions classified as C4 and C5 have significantly elevated \Ha velocity dispersion in comparison with that typical for their brightness that can be a consequence, among others, of a presence of an expanding superbubble resulting in a broadened or multi-component (resolved or unresolved) line profile. \revone{The expected triangle-shape peak (orange colour, C5) for such structures  is observed only in one region -- S2.} High velocity dispersion peaks are also observed in the regions S1, S3 and S4, but their triangle-shape is not obvious (probably because of overlapping). All these regions of high velocity dispersion are surrounded by C3 pixels that demonstrate a shallow change of the $\sigma(\mathrm{H}\alpha)$ expected for expanding superbubbles. 

\begin{figure*}
    \centering
    \includegraphics[width = 0.39\linewidth]{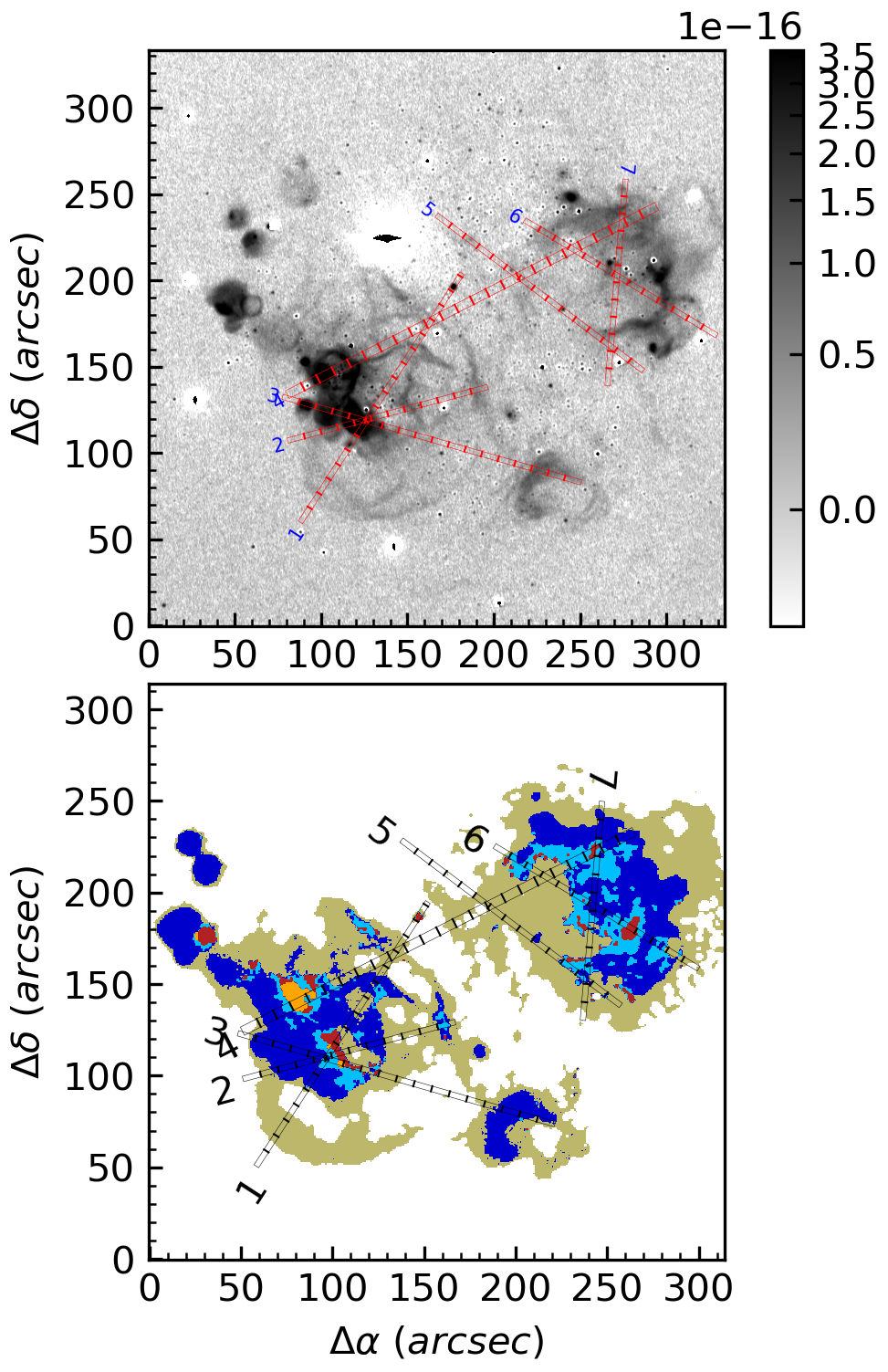}~\includegraphics[width = 0.61\linewidth]{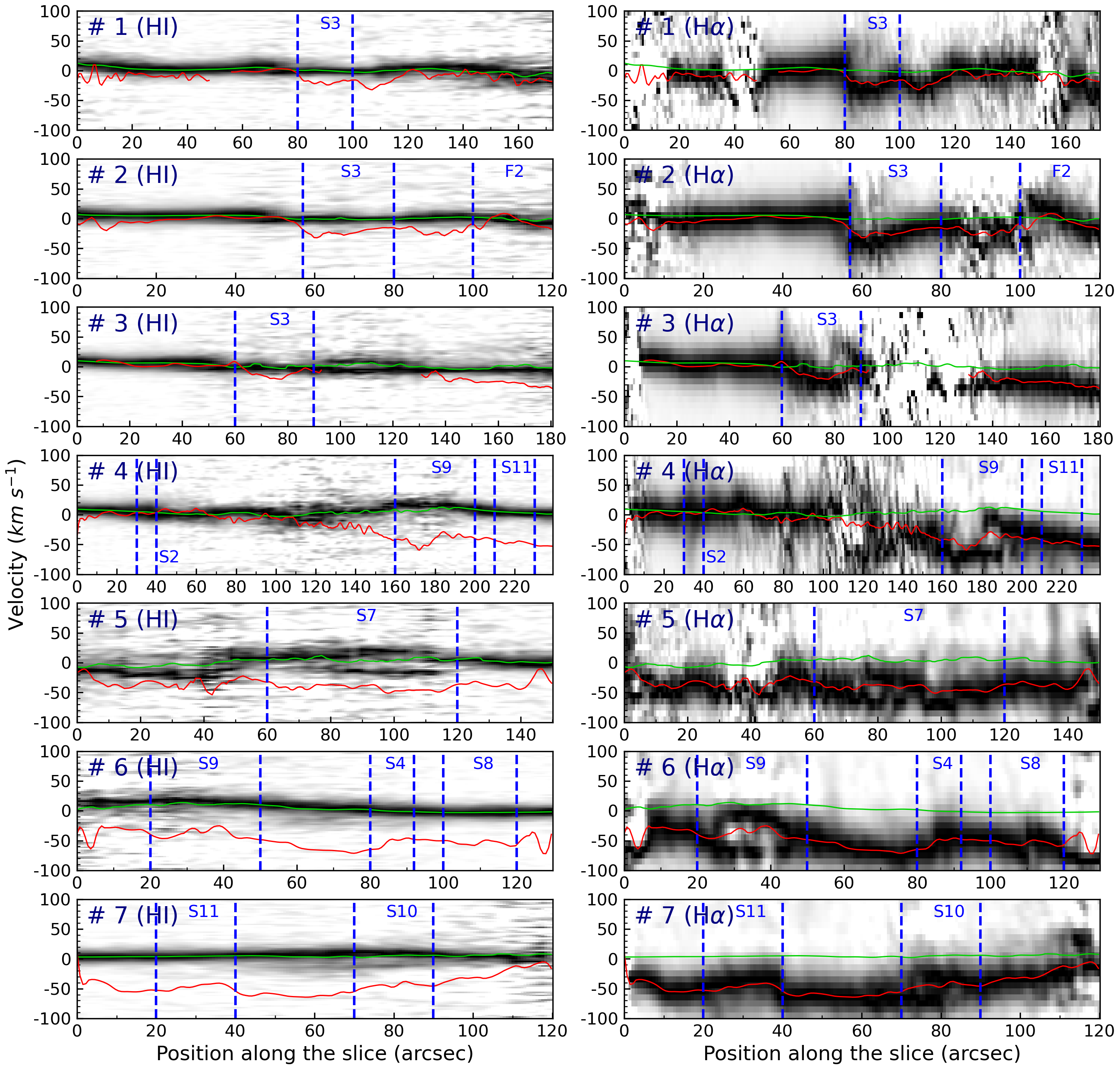}
    \caption{The KPNO \Ha image of the galaxy (top-left) and the $I-\sigma$ classification map (bottom-left). The  locations of the PV diagrams are superimposed. PV diagram names shown near the bar denote their position. Vertical ticks on each slice correspond to the $10\arcsec$ step.  PV diagrams are shown in the middle (\HI 21 cm) and right (H$\alpha$) panels. Red and green lines denote to the centre of the \Ha and \HI 21 cm line, respectively. \revone{The borders of the regions discussed in the text are shown by the blue dashed lines.} In order to highlight both faint and bright regions, the presented PV diagrams were normalized to the maximal brightness for each position along the slice.}
    \label{fig:pv_diagrams}
\end{figure*}

All the areas of high-velocity dispersion of ionized gas\revone{, that we identified on $I-\sigma$ diagram,} coincide with the regions S1--S4 selected as having a shell-like structure and clearly show a two-component line profile (see Fig.~\ref{fig:spectr_example}). We do not observe high velocity dispersion in the other superbubble candidates, though most of them were classified as C3 and thus still have velocity dispersion elevated in comparison with that normally distributed. 

\subsubsection{\revone{Position--velocity diagrams}}
\label{sec_pv}
In order to further emphasize a very perturbed kinematics of the ionized gas in Sextans~A, we show several position--velocity (PV) diagrams in Fig.~\ref{fig:pv_diagrams} in the \Ha and \HI 21~cm lines crossing the star-forming complexes B, C, D. So-called `velocity ellipse' (or sometimes only its part corresponding to either approaching or receding side of a bubble) is observed \revone{in the \Ha line} towards several our superbubble candidates: see positions 80--100 arcsec on PV~\#1, 60--90 on PV~\#3 for S3; 160--200 arcsec on PV~\#4 and 20--50 on PV~\#6 for S9; 70--90 arcsec on PV~\#7 for S10; 20--40 arcsec on PV~\#7 for S11. The overall `ellipse-like' change of the \Ha velocity is visible on the whole extent of PV~\#6 crossing the complex~C (S7) thus supporting the hypothesis of its probable expansion (or onset outflow). Another velocity ellipse is also visible \revone{in the \Ha line} on PV diagram \#1 (100--120 arcsec) and corresponds to the area in the complex~B outside the region S3 and encircled by the filamentary structure F4 in Fig~\ref{fig:spectr_example}, however we cannot identify its borders -- it does not have \revone{ clearly visible shell-like morphology in the \Ha line and is dominated by DIG in $I-\sigma$ diagram}.

The presented PV diagrams highlight two interesting features in the ionized gas kinematics in the complex~B. First is a steep variation of the \Ha velocity at the location where the region S3 is connected with the brightest \HII region ($\sim80$~arcsec on PV~\#1; $\sim60$~arcsec on PV~\#2). As follows from the $I-\sigma$ classification map, the highest velocity dispersion in this region is also observed there. This can again point to the outflowing gas from the \HII region driven by the feedback of massive stars, and we may expect to see other signatures of shocks \revone{(e.g. elevated line ratios of \SIIHa)} in this region. Another feature is a significant change in the \Ha velocity across the northern filament (part of it, marked as F2 in Fig.~\ref{fig:spectr_example}) visible on PV~\#2 ($\sim100-120$~arcsec). It is also well visible on \Ha velocity map (Fig.~\ref{fig:vel_map}e). Such velocity change and the general filamentary morphology of this region can also be a result of the impact of shock onto the ISM. However, we do not see clearly elevated velocity dispersion or underlying broad component in the filament. The exception is the northern part of it (F3 in Fig.~\ref{fig:spectr_example}) where the broad underlying component is clearly observed (see profile~\#3). 

From PV diagrams it is also noticeable that atomic \revone{gas} is significantly less perturbed than ionized gas in the studied star-forming complexes. The velocities in \HI 21~cm line are consistent with those in \Ha in the complex~B (PV~\#1--3), while a strong offset between \HI and \Ha is observed in the complex~C (PV~\#4--7; see also Sec.~\ref{sec:velocities}). In this complex~C the \HI kinematics is more perturbed and reveal several regions with multiple components (e.g. 20--50 and 70--120~arcsec on PV~\#5, 40--90~arcsec on PV~\#7), though there is no clear correlation with similar features in \Ha kinematics. As was already mentioned in Sec.~\ref{sec:velocities}, such discrepancy between \Ha and \HI is unusual and indicates that the emissions from ionized and atomic gas come from different regions along the line of sight. Blue-shifted \Ha emission can trace the outflowing gas in the complex~C, while the red-shifted or unperturbed \HI component can trace a far wall of the supergiant \HI shell surrounding star-forming part of the galaxy (see~Fig.~\ref{fig:map}).

\begin{table*}
    \caption{Properties of the identified expanding ionized superbubbles in Sextans~A.}
    \centering
    {\scriptsize
    \begin{tabular}{cccccccccccccc}
    \hline
         \#&RA(J2000)&DEC(J2000)&a, $^{\prime\prime}$&b, $^{\prime\prime}$&PA, $^\circ$&R, pc&\multirow{2}{.6cm}{\centering $t_\mathrm{kin}$,\\ $Myr$}& \multirow{2}{.95cm}{$V_\mathrm{exp}$\\ $km\ s^{-1}$}& \multirow{2}{1cm}{\centering$L_\mathrm{mech},$\\$ 10^{36} erg/s$}&\multirow{2}{1cm}{\centering$E_\mathrm{kin},$\\$ 10^{49} erg$}&\multirow{2}{.7cm}{$n_\mathrm{0}$,\\ $cm^{-3}$}&\multirow{2}{.6cm}{\centering $N$\\ $O5I$} \\
         &&&&&&&&&&& \\ 
    \hline
$S 1$&$152\degree47\arcmin21.2\arcsec$&$-4\degree41\arcmin34.2\arcsec$&    11&    10&$ 90$&$    35$&$  0.67$&$    31$&$    10.94$&$    23.10$&$  0.76$&$   13$ \\ 
$S 2$&$152\degree46\arcmin31.5\arcsec$&$-4\degree42\arcmin08.6\arcsec$&    19&    12&$ 58$&$    48$&$  1.14$&$    25$&$    12.72$&$    45.68$&$  0.87$&$   15$ \\ 
$S 3$&$152\degree46\arcmin01.7\arcsec$&$-4\degree42\arcmin35.5\arcsec$&    34&    20&$ 53$&$    82$&$  2.58$&$    19$&$     7.07$&$    57.80$&$  0.37$&$    8$ \\ 
$S 4$&$152\degree43\arcmin32.0\arcsec$&$-4\degree41\arcmin32.0\arcsec$&    24&    16&$140$&$    62$&$  1.69$&$    22$&$    10.37$&$    55.39$&$  0.63$&$   12$ \\ 
$S 5$&$152\degree45\arcmin59.8\arcsec$&$-4\degree42\arcmin41.0\arcsec$&   120&    89&$140$&$   331$&$ -$&$    -$&$   -$&$-$&$-$&$ -$ \\ 
$S 6$&$152\degree44\arcmin21.0\arcsec$&$-4\degree43\arcmin25.5\arcsec$&    41&    33&$ 58$&$   118$&$  5.45$&$    13$&$     3.66$&$    63.05$&$  0.30$&$    4$ \\ 
$S 7$&$152\degree44\arcmin05.6\arcsec$&$-4\degree41\arcmin23.9\arcsec$&    90&    75&$120$&$   262$&$  4.62$&$    34$&$   417.31$&$  6091.25$&$  0.39$&$  499$ \\ 
$S 8$&$152\degree43\arcmin13.7\arcsec$&$-4\degree41\arcmin45.1\arcsec$&    34&    23&$140$&$    90$&$  2.24$&$    24$&$    41.54$&$   294.28$&$  0.93$&$   49$ \\ 
$S 9$&$152\degree44\arcmin14.9\arcsec$&$-4\degree40\arcmin59.7\arcsec$&    24&    18&$115$&$    67$&$  1.50$&$    27$&$    10.49$&$    49.76$&$  0.30$&$   12$ \\ 
$S10$&$152\degree43\arcmin56.3\arcsec$&$-4\degree41\arcmin38.8\arcsec$&    20&    17&$ 87$&$    59$&$  1.50$&$    24$&$     6.24$&$    29.53$&$  0.34$&$    7$ \\ 
$S11$&$152\degree43\arcmin47.1\arcsec$&$-4\degree40\arcmin51.4\arcsec$&    17&    16&$139$&$    53$&$  1.34$&$    24$&$    11.11$&$    46.92$&$  0.71$&$   13$ \\   \hline
    \end{tabular}}
    \label{tab:bubble_params}
\end{table*}

\subsubsection{\revone{Expanding superbubbles}}
\label{sec_halpha_sum}
Summing up, from the kinematical analysis presented in this section and above, we were able to identify 10 expanding ionized superbubbles in the star-forming complexes~B, C and D. Their diameters vary from 70 to 500~pc. One additional region (S5) also has a shell-like morphology and is encircled by two extended filaments having opposite residual velocities. The northern filament has signs of shock wave impacts \revone{(underlying broad component; see Fig.~\ref{fig:spectr_example}, profile~\#3)} in its \Ha kinematics. While it is not clear if S5 is a single structure or a result of a projection, it can be a relatively old stalled supershell driven by previous supernovae explosions in the complex~B -- this scenario is in agreement with the estimates of the age of the complex ($\sim 200$~Myr according to \citealt{Camacho2016}).

From the decomposition of the \Ha line profiles in the identified superbubbles (Fig.~\ref{fig:spectr_example}), we can estimate their expansion velocities. For the large superbubbles S6 and S7 we considered the maximal value of the residual velocities in Fig.~\ref{fig:vel_map}e as a proxy of their expansion velocities (given that velocities of their bright rims are close to zero). We cannot estimate the expansion velocity of the supershell S5, and as was mentioned above - we are not even sure that this is a single structure. Assuming the \citet{Weaver1977} model of expanding superbubbles, we estimated their kinematical age as

\begin{equation}
    t_\mathrm{kin} = 0.6R/V_\mathrm{exp},
\end{equation}
where $R = \sqrt{R_\mathrm{a}R_\mathrm{b}}/2$ - effective radius of the superbubble in pc; $R_\mathrm{a}$ and $R_\mathrm{b}$ are the major and minor axes (in parsecs) of the ellipse encircling the region, respectively; $V_\mathrm{exp}$ is the measured expansion velocity in $\kms$. Our findings are summarized in Table~\ref{tab:bubble_params}, which contains coordinates of the centres of each superbubble, their measured major and minor axes ($a$ and $b$), positional angle (PA), effective radius ($R$), 
kinematical age ($t_\mathrm{kin}$, expansion velocity ($V_\mathrm{exp}$), mechanical luminosity ($L_\mathrm{mech}$) and total kinetic energy $E_\mathrm{kin}$ according to the \cite{Weaver1977} model (see Section~\ref{sec_an_energy}), volumetric atomic number density ($n_\mathrm{0}$), and the equivalent number of O5I stars ($N$) necessary to form the superbubble as result of pre-supernova feedback (see Section~\ref{sec_an_energy}).

\subsection{Gas excitation state.}
\label{sec:ion_state}

\begin{figure}
    \centering
    \includegraphics[width = \linewidth]{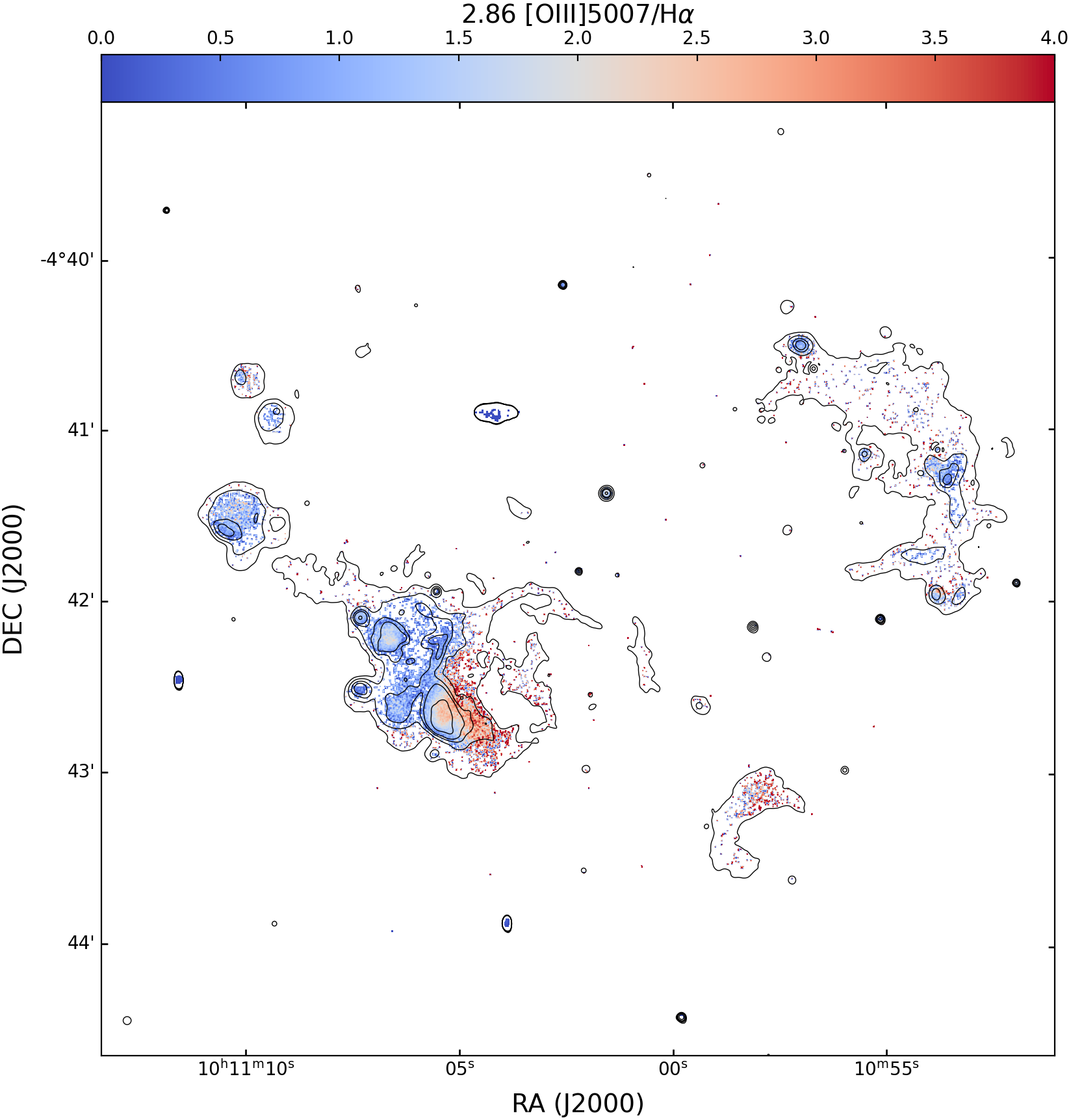}
\caption{[O~\textsc{iii}]~5007\AA/$H\beta$ emission lines ratio map for Sextans~A. The map in \Hb line was obtained from \Ha map assuming a theoretical ratio of H$\alpha$/H$\beta = 2.86$. All pixels with $\mathrm{S/N} < 2$ in \OIII\, lines or $\mathrm{S/N} < 7$ in \Ha were masked. Contours are the lines of a constant \Ha surface brightness.}
    \label{fig:map_oiii}
\end{figure}

\begin{figure}
    \centering
    \includegraphics[width = \linewidth]{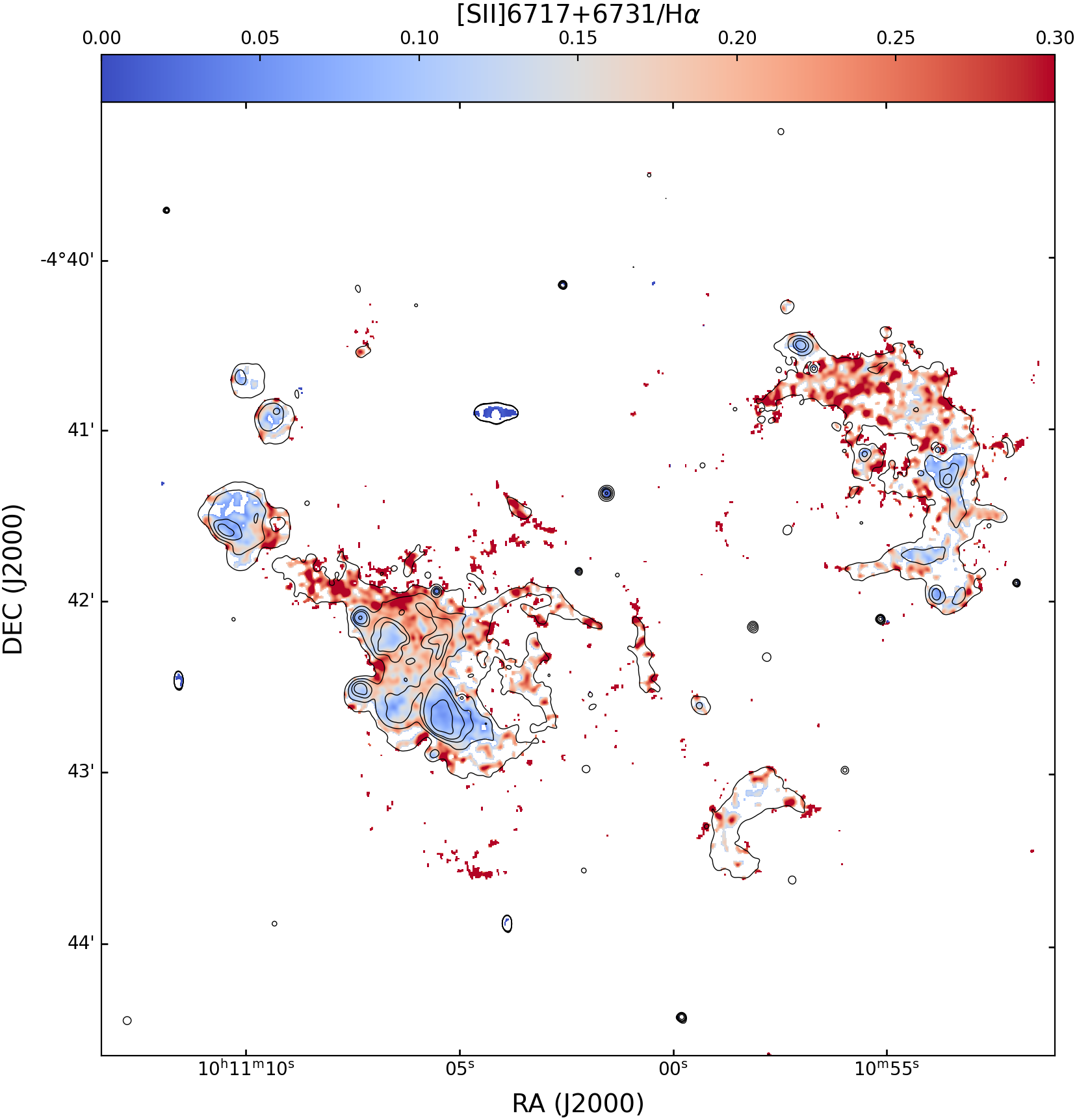}
    \caption{[S~\textsc{ii}]~6717,6731\AA/$H\alpha$ emission lines ratio map for Sextans~A. All pixels with $\mathrm{S/N} < 3$ in \SII\, lines or $\mathrm{S/N} < 5$ in \Ha were masked. Contours are the lines of a constant \Ha surface brightness.}
    \label{fig:map_sii}
\end{figure}

Strong emission line ratios are common tools for diagnostics of the gas ionization state. The most important line ratios are \OIIIHb, \NIIHa\, and \SIIHa, which are used in the so-called BPT diagram (named after \citealt{BPT}) and its variations \citep{Veilleux1987}. We used the \OIIIHb\, and \SIIHa\, emission line ratios to trace the spatial variations of the gas ionization state in Sextans~A. For that we rely on our narrow-band images in \SII\, and \OIII\, lines obtained with MaNGaL, and on the archival \Ha images (see Sections~\ref{sec_obs_figs} and \ref{sec_obs_other}). Before the analysis, all the images were convolved to the same angular resolution ($FWHM = 2.4^{\prime\prime}$) and aligned to the same grid. Since we do not have \Hb line map, we used a theoretical value of a Balmer decrement ($\mathrm{H\alpha/H\beta = 2.86}$ see e.g. \citealt{Osterbrock2006}) and produced \revone{\Hb map from the \Ha map assuming theoretical ratio of H$\alpha$/H$\beta = 2.86$.} Foreground extinction in direction to the Sextans~A is very small ($A_\mathrm{V} = 0.122$ \citealt{Schlafly2011}) and does not affect the considered line ratios.

The maps of \OIIIHb\, and \SIIHa\, emission lines ratios are shown in Figs.~\ref{fig:map_oiii} and \ref{fig:map_sii}, respectively. Though \OIII\, is not deep enough to trace the line ratios in all regions of our interest, including filaments surrounding the complex B, it gives several \revone{results}. The most prominent feature is a strong gradient of \OIIIHb\, from the brightest \HII region in the complex~B towards the lower density environment encircled by the northern filament; this area of high \OIIIHb\ coincides with the superbubble S3. This supports our hypothesis about the outflow of the ionized gas in this region (see further discussion in Sec.~\ref{sec:disc_outflow}). Other bright \HII regions have much lower \OIIIHb\, ratio and ionization structure typical for \HII regions (with the higher ratio of \OIIIHb\, in their central parts, see e.g. \citealt{Pellegrini2012}). Another region with a relatively high \OIIIHb\ ratio is the northern part of the complex~D. This area corresponds to one of the youngest OB stars in the galaxy discovered by \cite{Garcia2019}. We will consider this object in more detail in Sec.~\ref{sec:disc_star}.

Because \SIIHa\, line ratio is sensitive to the shock waves \citep[e.g.][]{Allen2008}, the criteria of \SIIHa$>0.4$ is commonly used to search for the supernovae remnants in nearby galaxies \citep[e.g.][]{Blair1981, Blair2004}. This threshold is however valid for solar metallicity, while the typical values of \SIIHa\, produced by shocks are significantly lower in the low-metallicity environment like in Sextans~A. 
According to the models from \cite{Alarie2019}, at $Z \sim 0.13 Z_\mathrm\odot$ shocks should typically produce \SIIHa~$\gtrsim0.25$ (and \SIIHa~$\gtrsim0.15$ at $Z \sim 0.07Z_\mathrm\odot$). As follows from the Fig~\ref{fig:map_sii}, all regions bright in \Ha have \SIIHa\, line ratio below this threshold with slightly increased values between the bright \HII regions in the complex~B that is probably a result of the high contribution of DIG. Meanwhile, one can see hints higher \SIIHa\, towards the filaments (both northern and southern) surrounding the complex~B, in the north part of the complex~C (towards the superbubble S11 where we observed a broad underlying component, see Fig.~\ref{fig:spectr_example}) and in the small superbubble S1. We note however that S/N in these areas is insufficient for certain conclusions on that from narrow-band maps analysis. 

\begin{figure*}
    \centering
    \includegraphics[width = \textwidth]{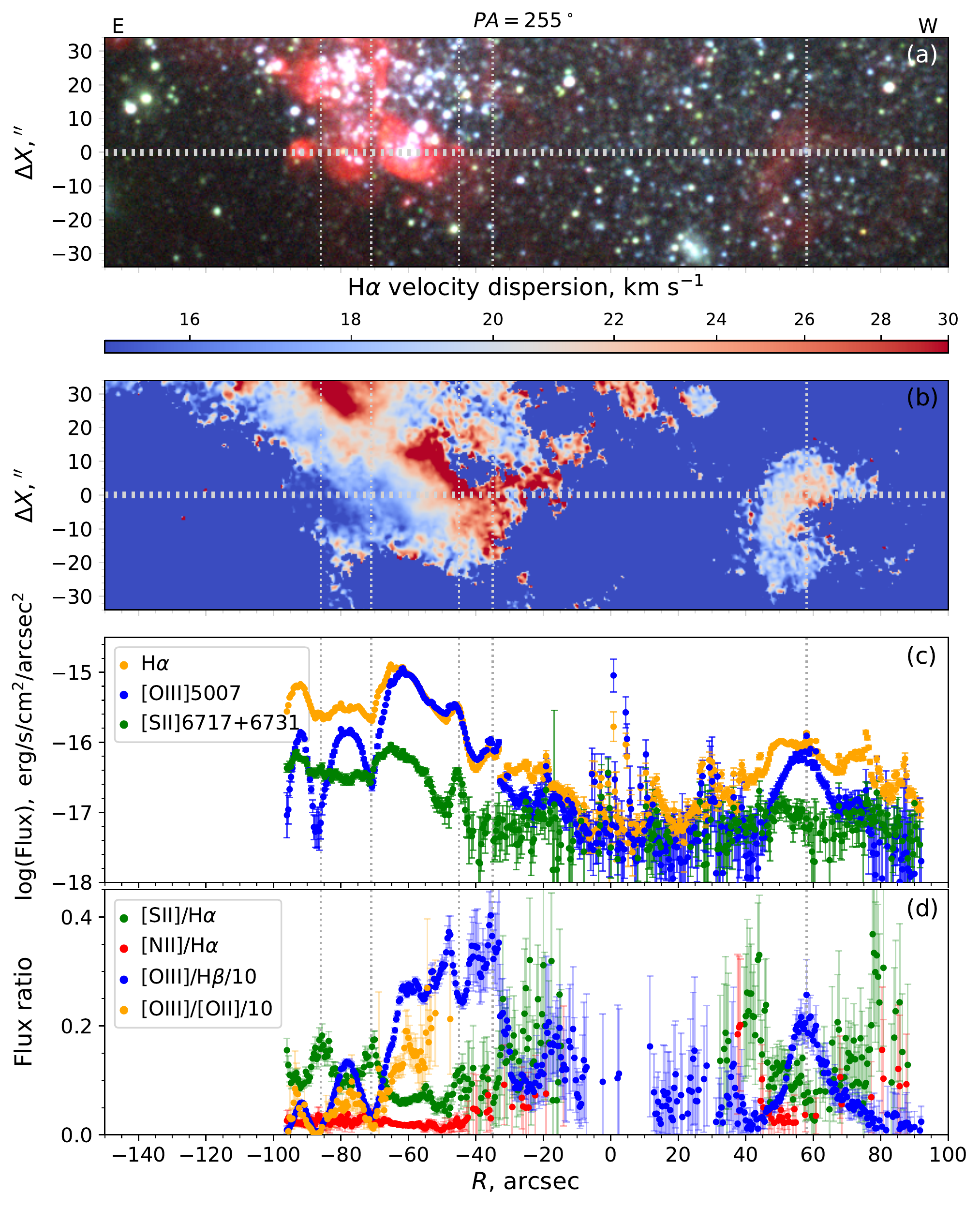}
    \caption{Distribution of the emission line fluxes (panel c) and their ratios (panel d) along the slit crossed the southern part of the Sextans~A. Localization of the slit is shown by horizontal line in panels (a) and (b). Panel (a) shows the optical image in \Ha and continuum (\Ha in the red channel, $R$ and $V$-band images in green and blue channels, respectively). Panel (b) demonstrates the distribution of the \Ha velocity dispersion measured from FPI data.}
    \label{fig:line_spectra}
\end{figure*}

Fig.~\ref{fig:line_spectra} shows the localization of the SCORPIO-2 long-slit spectrum crossing the brightest \HII region in the complex~B and \revone{also young massive star S004 (according to the list  \citealt{Lorenzo2022}) in the complex~D, which we will discuss later in Sec.~\ref{sec:disc_star}.} We present there the distribution of several bright emission line fluxes and their ratios along the slit. Their distribution is consistent with what was found from the narrow-band images. Highest \OIIIHb\, line ratios is observed at the edge of the brightest \HII regions (in the superbubble S3; \revone{$-50...-35$ arcsec in Fig.~\ref{fig:line_spectra}}) suggesting that it is filled by hot gas. Also [O~\textsc{iii}]/[O~\textsc{ii}] ratio, a proxy of ionization parameter, \revone{increases towards the interior of the S3 superbubble}. We do not see obvious signatures of shocks in \SIIHa\, line ratio, however, some contribution can take place at the edge of the superbubble S3 ($-30$...$-10$ arcsec, where both \SIIHa\, and \OIIIHb\, are elevated), and also in the point-like emission source at $\sim -45$~arcsec (as it shows a prominent peak in \SII\, line together with elevated \OIIIHb, and also has an underlying broad component in its \Ha line, see profile~\#11 in Fig.~\ref{fig:spectr_example}). We note that \SIIHa\, shows local peaks also at $\sim -90$, $-70$, 40 and 80 arcsec in the long-slit spectrum -- they correspond to the low brightness regions and thus are probably associated with DIG. 

\begin{figure}
    \centering
    \includegraphics[width=\linewidth]{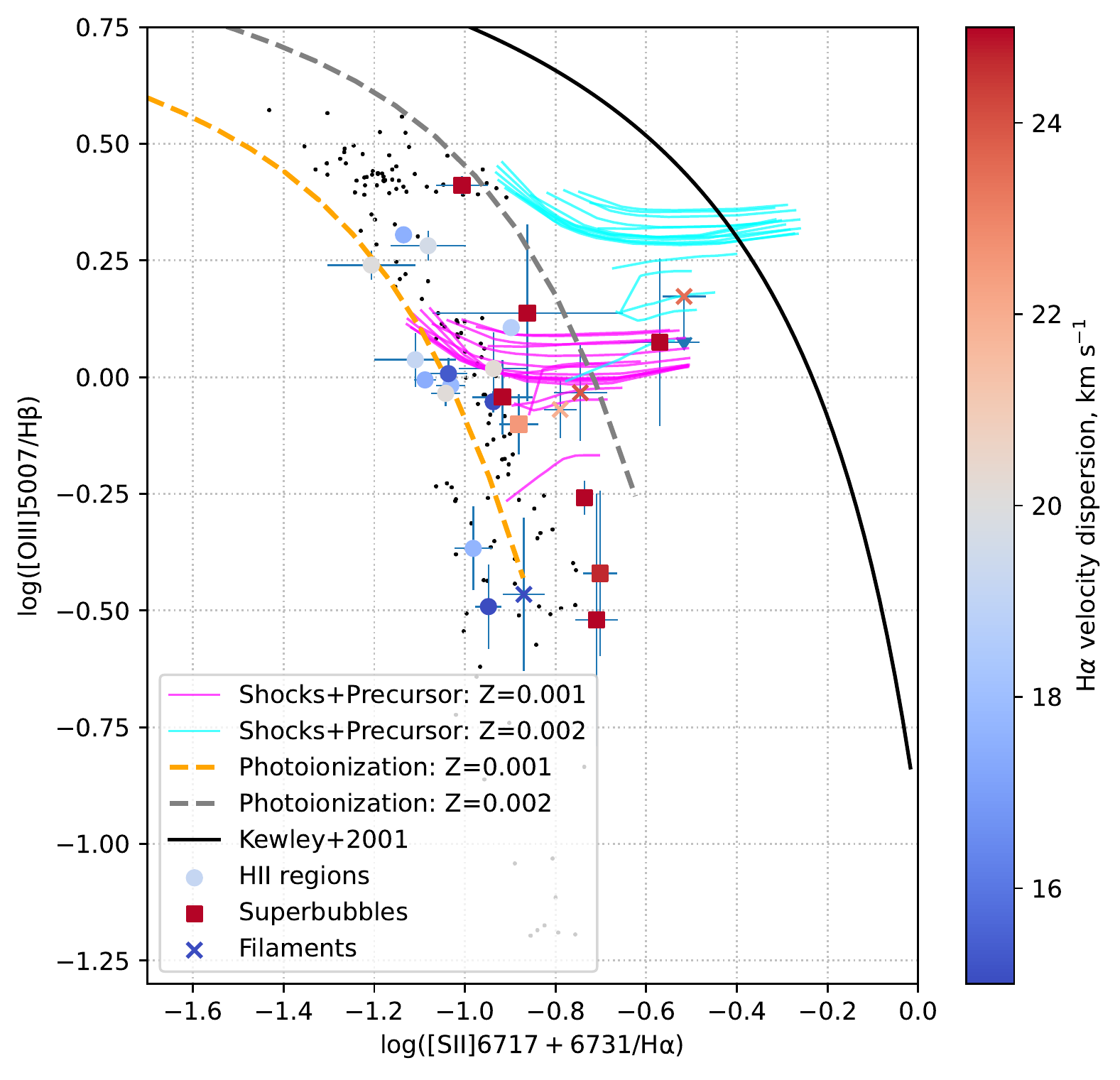}
    \caption{Diagnostic diagram showing \OIIIHb\, vs \SIIHa\, for different types of nebulae (bright \HII regions, superbubbles, filaments, shown by different symbols) constructed by integration of the flux in narrow-band images in apertures shown in Fig.~\ref{fig:spectr_example}. Colour shows the \Ha velocity dispersion measured from the line profiles integrated over the same apertures. Small black dots show the values obtain for bright pixels ($F\mathrm{(H\alpha) > 10^{-16}\ erg\ s^{-1}cm^{-2} arcsec^{-2}}$) as a reference. Black solid line is the `maximum starburst line' from \citet{Kewley2001}. Dashed lines correspond to the photoionization models from \citet{Gutkin2016} for two metallicities. Cyan and magenta lines show grids for low-metallicity shock models from \citet{Alarie2019} for similar values of metallicities.}
    \label{fig:bpt_img}
\end{figure}

\begin{figure*}
    \centering
    \includegraphics[width=\linewidth]{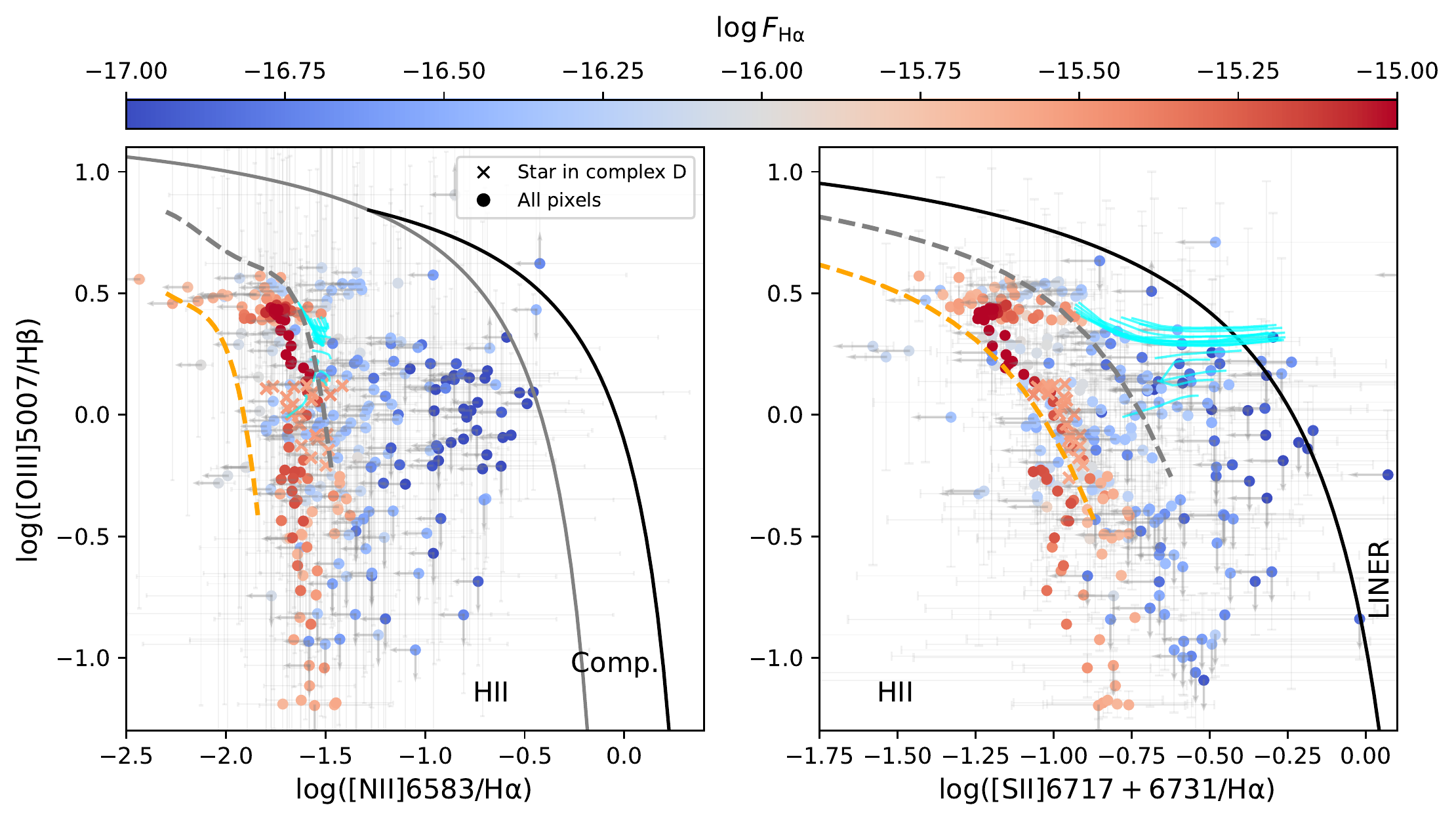}
    \caption{Diagnostic BPT diagrams shown for each pixel along the long slit spectrum presented in Fig.~\ref{fig:line_spectra}. Colour denotes the \Ha surface brightness. Crosses mark the region 55-65~arcsec along the slit corresponding to the young OB star in the complex~D and its vicinity (see~Sec.~\ref{sec:disc_star}).  The designation of the lines is the same as in Fig.~\ref{fig:bpt_img}. An additional grey line corresponds to the \citet{Kauffmann2003} line separating the area occupied by pure \HII regions from that with the composite mechanism of ionization.}
    \label{fig:bpt_ls}
\end{figure*}

As follows from Figs.~\ref{fig:map_oiii},~\ref{fig:map_sii}, our narrow-band images are not deep enough for the analysis of the spatial distribution of the line ratios in faint emission structures. To increase S/N in these structures, we integrated the flux in each of the involved emission lines in the apertures shown in Fig.~\ref{fig:spectr_example}. Namely, we considered all superbubbles (except largest S5, S6 and S7), parts of filaments denoted as F1--F4, and several bright \HII regions \revone{for comparison} (marked by green circles in Fig.~\ref{fig:spectr_example}). The BPT diagram corresponding to such measured values for these structures is shown in Fig.~\ref{fig:bpt_img}. Small points demonstrate the values extracted from the individual pixels along the long-slit spectrum and having $F\mathrm{(H\alpha) > 10^{-16} erg\ s^{-1}\ cm^{-2}\ arcsec^{-2}}$. Good agreement between these values validates our estimates obtained from narrow-band images. The BPT diagrams for all pixels from the long-slit spectrum are shown in Fig.~\ref{fig:bpt_ls} as well. Several results can be emphasized from these diagrams:
\begin{itemize}
    \item Almost all bright \HII regions follow the line corresponding to the photoionization model from \cite{Gutkin2016} for metallicity $Z = 0.001 (\sim0.07Z_\odot)$. At the same time, most of the superbubbles and filaments depart from this model and show line ratios that are more consistent with the shock models from \citet{Alarie2019} having the same metallicity prescription as the \cite{Gutkin2016} models, or with the photoionization model for a twice larger metallicity. We note that in Fig.~\ref{fig:bpt_ls}, photoionization models for different metallicities \revone{have} different line ratios. Despite that, all bright regions there can be described by pure photoionization, while other mechanisms (including shocks) significantly contribute to the ionization of fainter regions (including shells and filaments). 
    \item Position of the different regions correlate with their \Ha velocity dispersion \revone{(see Fig.~\ref{fig:bpt_img})}. Such a `BPT--$\sigma$' trend with velocity dispersion is often observed in the regions where the shocks contribute significantly to the overall ionization balance \citep[e.g.][]{Oparin2018, DAgostino2019}, though this can be also a consequence of higher velocity dispersion in the diffuse ionized gas. \citep[e.g.][]{MoiseevKlypin2015, DellaBruna2020}. In our case, the higher velocity dispersion of superbubbles is rather a consequence of their expansion (see Sec.~\ref{sec:superbubbles}), while shocks can significantly contribute to the excitation (and higher velocity dispersion) for the northern filament. \revone{The line ratios of its parts (especially of F3) agree well with the grids of low-metallicity shock models (see Fig.~\ref{fig:bpt_img})}. 
\end{itemize}

We note also that the classical criteria for separating the different ionization mechanisms on BPT diagrams (see the demarcation lines from \citealt{Kewley2001} and \citealt{Kauffmann2003}) were developed for solar metallicity and thus not appropriate for low-metallicity objects like Sextans~A. At low metallicity, shocks can produce line ratios very similar to that for pure photoionization, and relative variations of the line ratios are more important than considering these classical criteria.

\section{Discussion}
\label{sec:discussion}

\subsection{Energy balance between the massive stars and ionized gas.} \label{sec_an_energy}

\begin{figure}
    \includegraphics[width = \linewidth]{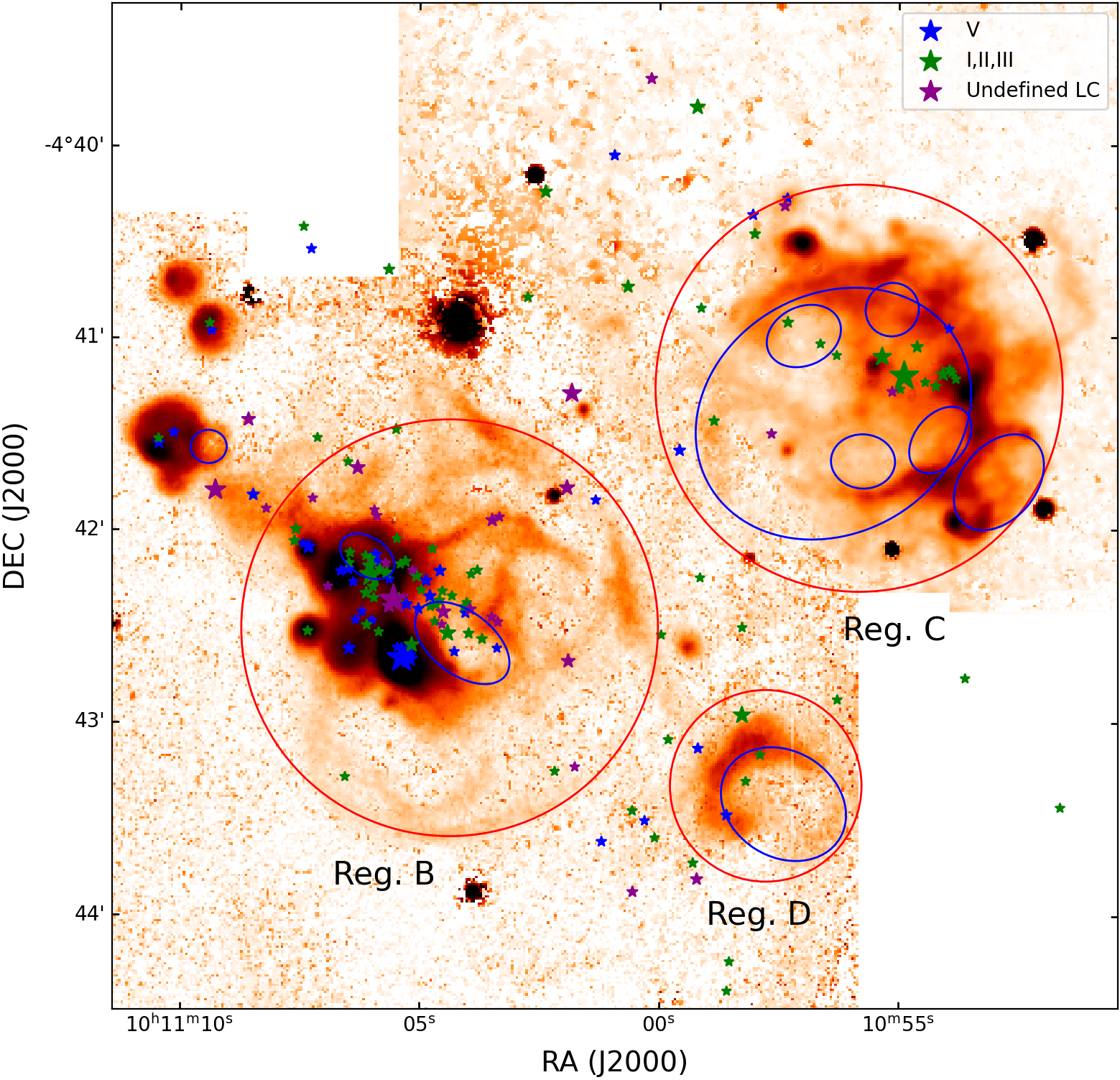}
    \caption{\Ha map of the Sextans~A according to the scanning FPI observations. Asterisks correspond to the locations of OB stars identified by \citep{Lorenzo2022}. Different colours correspond to different luminosity classes. Blue ellipses mark the positions of the identified superbubbles.}
    \label{fig:stars}
\end{figure}

Massive stars in star-forming complexes of Sextans~A are the most probable sources of the ionization of the ISM, and their feedback is likely drives the kinematical features described in previous Sections. It is still worth quantifying, however, whether \revone{massive stars provide enough energy to blow out cavities in gas} and how efficiently feedback regulates the ISM. Thanks to the proximity of Sextans~A, the individual young OB stars can be identified in the \HST images. \citet{Lorenzo2022} published recently the catalogue of such spectroscopically confirmed massive stars among those UV sources identified by \cite{Bianchi2012} in the \HST data. We use this catalogue and the spectral classification of the stars in their work (see Fig.~\ref{fig:stars}).

\subsubsection{Ionization balance and escape fraction of ionizing photons}

Relying on the spectral types of each massive star in the complexes B, C and D, we can estimate the total amount of the hydrogen ionizing photons ($Q^0_\mathrm{stars}$) and compare this with that required to produce the observed \Ha flux (corrected for foreground reddening) in each region ($Q^0_\mathrm{H\alpha}$). For that we rely on the models of OB stars of different spectral and luminosity classes from \citet{Martins2005} presented for different metallicities. Since there are many early-type O stars in the galaxy, the contribution of the late-type stars (older than O9.5) is insignificant \revone{(see e.g. fig. 12 in \citealt{Ramachandran2019})} and we do not consider them in the calculations. Assuming case B recombination, the value of $Q^0_\mathrm{H\alpha}$ is related to the \Ha luminosity as \citep{Osterbrock2006}:

\be \frac{L(\mathrm{H\alpha)}}{h\nu_\mathrm{H\alpha}} \simeq \frac{\alpha^\mathrm{eff}_\mathrm{H\alpha}}{\alpha_{\mathrm{B}}}Q^0_\mathrm{H\alpha} \simeq 0.45 Q^0_\mathrm{H\alpha}, \ee
where the total recombination coefficient of hydrogen $\alpha_B \simeq 2.59 \times 10^{-13}\ \text{cm}^3\ \text{s}^{-1}$, the effective recombination coefficient in \Ha is $\alpha^\mathrm{eff}_\mathrm{H\alpha} \simeq 1.17 \times 10^{-13}\ \text{cm}^3\ \text{s}^{-1}$ at $T_{\rm{e}} = 10\,000\ \mathrm{K}$. The obtained values of $Q^0_\mathrm{stars}$ and $Q^0_\mathrm{H\alpha}$ are given in Table~\ref{tab:ion}. For all three complexes O stars from the catalogue by \cite{Lorenzo2022} produce sufficient amount of ionized quanta to ionize the ISM. Moreover, about 55--70 per cent of the ionizing photons should escape from the \HII complexes (even from the complex~D with the small number of O stars) and potentially produce the bulk of DIG; some of these quanta might also escape from the galaxy. 

\begin{table}
    \centering
    \caption{Ionization quanta balance between massive stars and ISM.}
    \begin{tabular}{ccccc}
         \hline
         Complex & $Q^0_\mathrm{H\alpha}$ & $Q^0_\mathrm{stars}$ & $f_{esc}$ \\
         \hline
        B & $  1.08 \cdot 10^{50} $& $  2.42 \cdot 10^{50}$ & $0.55$ \\
        C & $  7.57 \cdot 10^{49} $& $  1.78 \cdot 10^{50}$ & $0.58$ \\
        D & $  1.18 \cdot 10^{49} $& $  3.75 \cdot 10^{49}$ & $0.69$ \\

         \hline
    \end{tabular}
    \label{tab:ion}
\end{table}

High values of escape fraction of ionizing photons were measured for \HII regions in several nearby galaxies \revone{\citep[see, e.g.,][]{Oey1997,Egorov2018,Weilbacher2018, McLeod2019,Choi2020, DellaBruna2021, Egorov2021, Ramambason2022}. Most of these studies usually derive the escape fraction to be $\sim 40-60$ per cent, and thus our estimates for Sextans~A are on the upper end.} According to \cite{Niederhofer2016}, the values of $Q^0_\mathrm{stars}$ bear large uncertainties and can be affected by the choice of a particular model up to a factor of 2. Even if the values of $Q^0_\mathrm{stars}$ in Table~\ref{tab:ion} are overestimated by a factor of two, the resulting escape fraction from the star-forming complexes in Sextans~A should remain significant. 

\subsubsection{Are the superbubbles created by supernovae?}
\label{sec:supernovae}

In previous Sections we described the properties of 11 shell-like structures of ionized gas and demonstrated that 10 of them are probably expanding superbubbles. From the measured observational properties we can estimate the required mechanical luminosity $L_\mathrm{mech}$ to blow-out these superbubbles, and the total kinetic energy $E_\mathrm{kin}$ accumulated throughout the evolution of the superbubble. 
The evolution of a superbubble around OB associations driven by winds and SNe can be described by the following equations \citep{MacLow1988}:

\be R_\mathrm{s}(t) = 66(\frac{L_{38}}{n_0})^{1/5}t_6^{3/5}; \ee
\be V_\mathrm{s}(t) = 3R_s/5t = 38.6(\frac{L_{38}}{n_0})^{1/5}t_6^{-2/5}, \ee
where $R_\mathrm{s}(t)$ is time-dependent radius of the shell in pc, $V_\mathrm{s}(t)$ is expansion velocity (in $\kms$), $L_{38} = L_\mathrm{mech}/10^{38}$ $\mathrm{erg\ s}^{-1}$, $n_0 = n_\mathrm{H}+n_\mathrm{He} = \mu n_\mathrm{H}$ is atomic number density in $\mathrm{cm}^{-3}$, $\mu = 1.4$, $t_6$ is age of the superbubble (in Myr). Thus, the required mechanical luminosity can be estimated from the measured values in Table~\ref{tab:bubble_params} for each superbubble as
\be L_\mathrm{mech} = 3.99\cdot10^{29}{{n_0}R^2 V_\mathrm{exp}^{3}}. \ee

The atomic number density $n_0$ was estimated from the column density $N_\mathrm{HI}$ of neutral hydrogen assuming the scale height of the \HI disc of the galaxy $h = 438$~pc \citep{Stilp2013}. The measured column density $N_\mathrm{HI}$ is related to the average volume density $n_\mathrm{HI}$ as

\begin{equation}
    N_\mathrm{HI} = \int^{+\infty}_{-\infty}n_\mathrm{HI}\exp{\frac{-z^2}{2h^2}}dz = \sqrt{2\pi}hn_\mathrm{HI}.
\end{equation}
The inclination of the galactic disc increases the size the \HI column along the line of sight as $h/\cos{i}$, where $i = 34^\circ$ (see Table~\ref{tab:SexA}. Given that, we derive atomic number density for each superbubble as 
\begin{equation}
n_{0} = \frac{\mu N_\mathrm{HI} \cos{i} }{\sqrt{2\pi}h}.
\end{equation}
All derived values of $n_0$, $L_\mathrm{mech}$ and $E_\mathrm{kin} = L_\mathrm{mech}\times t_{kin}$ for 10 superbubbles are given in Table~\ref{tab:bubble_params}.

As follows from Table~\ref{tab:bubble_params}, the required energy to drive the expansion of almost all the identified superbubbles is on the order of $0.1-0.5\times10^{51}$~erg. Due to radiative cooling during their evolution, superbubbles can retain only about 30~per cent of the energy initially injected by stars \citep[e.g.][]{Sharma2014, Vasiliev2015}, which is not considered in our estimate, and this fraction can be different in low metallicity environment (Egorov et al., in prep.). Given this and that the energy of typical SNe type II is $E_\mathrm{SN} \sim 10^{51}$~erg, the explosion of $\sim1-2$ SNe can be sufficient to create most of the expanding superbubbles considered in this paper. 

Superbubble S1 is the youngest on the list. According to our estimates, it can be produced by 1--2 supernovae. Given its location at the edge of the bright isolated \HII regions, very round shape, presence of the broad component in its line profile (\#16 in Fig.~\ref{fig:spectr_example}) and locally elevated \SIIHa\, line ratio, we suggest this is a supernova remnant. A similar scenario is also probable for the superbubble S2 (demonstrating clear peak on $I-\sigma$ diagram, see Fig.~\ref{fig:i-sigma}), although the elevation of the \SIIHa\, in Fig.~\ref{fig:map_sii} is less obvious. Finally, S9 is probably driven by SN explosions because of its position on the BPT diagram suggests a significant contribution of shocks (see Fig.~\ref{fig:bpt_img}). Other superbubbles requiring small energy input also \revone{can appear as a result of SN explosion}, but they do not exhibit conclusive evidence in the considered data. 
Superbubble S7 surrounding the complex C requires significantly more supernovae to drive its expansion -- about $100-140$, if accounting for the radiative losses. We will discuss it later in Sec.~\ref{sec:disc_outflow}.

Thus, all considered superbubbles (but S7) could be created by a few supernovae. Although the age of these superbubbles is lower than 2.5~Myr in general -- the stars do have not enough time to evolve and explode except the most massive ones. However, if the pre-SN feedback contributes insignificantly to the energy balance, then the derived kinematic age can trace the period since the first (or the only) supernova explosion, and not since the onset of the stellar feedback. Nevertheless, it is worth checking whether pre-SN alone can provide sufficient energy to create these superbubbles and drive their expansion. 

\subsubsection{How significant is pre-supernovae feedback?}

At solar metallicity, massive O stars during their evolution can inject to the ISM near the same amount of energy \revone{through the wind} as at the moment of their explosion. At low metallicities the relative contribution of the winds \revone{to the total mechanical energy input} changes. Here we estimate how many O stars we need to create the identified superbubbles and support their expansion with the winds only. 

Given the value of $L_\mathrm{mech}$ required to drive the expansion of the superbubbles (see Table~\ref{tab:bubble_params}), we can estimate an effective number of O5I stars necessary to provide desired value of mechanical luminosity. For that we considered the parameters of the model OB\#14 from \citet{Smith2002} for metallicity Z = $0.05Z_\odot$ having mass-loss rate $\log(\dot{M}) = -5.96$ $\mathrm{M}_\odot\ \mathrm{year}^{-1}$ and wind velocity $v_\infty = 1550\ \kms$ and calculated wind luminosity for a single star $L_\mathrm{w} = \dot{M}v_\infty^2/2 = 8.35\cdot 10^{35}\ \mathrm{erg\ s^{-1}}$. From that we estimated $N(O5I)$ -- number of stars required to blow out each superbubble (given in Table~\ref{tab:bubble_params}).

We obtained that for most of the superbubbles 3--10 O5I stars can provide sufficient wind luminosity for driving their expansion. From comparison with the localization of the young massive stars (Fig.~\ref{fig:stars}) one can see that they are not always present in the superbubble. We cannot exclude however the incompleteness of the considered list of massive stars because we rely only on the spectrally confirmed objects from \cite{Lorenzo2022}.

We note that we consider here supergiants and not the main sequence stars because they are very abundant in the catalogue presented (84 supergiants of 116 total detected OB stars with known spectral type and luminosity class) by \cite{Lorenzo2022} (Fig.~\ref{fig:stars}). Considering the main sequence stars as the main contributors to pre-SN stellar feedback will increase the required effective number of stars by a factor of 13 -- in this case winds will fail to explain the observed properties of superbubbles.


\begin{table}
\caption{Comparison of the relative contribution of different pressure terms for superbubbles in Sextans~A.}
    \label{tab:pressure_energy}
    \centering
    \begin{tabular}{ccccc}
    \hline
         \#&\multirow{2}{1.2cm}{\centering $P_\mathrm{b}/n_{0},$ \\ $10^{-12} erg$}
         &\multirow{2}{1.3cm}{$P_\mathrm{rad}/n_{0},$ \\ $10^{-12} erg$}
         &\multirow{2}{1.7cm}{\centering $P_\mathrm{therm}/n_{0},$ \\ $10^{-12} erg$}&$100\times\frac{P_\mathrm{therm}}{P_\mathrm{b}}$ \\ &&&& \\
         \hline
 1&$ 16.06$&$  0.15$&$  2.76$&$ 17.2$ \\
 2&$ 10.59$&$  0.44$&$  2.76$&$ 26.1$ \\
 3&$  6.08$&$  0.21$&$  2.76$&$ 45.4$ \\
 4&$  8.05$&$  0.18$&$  2.76$&$ 34.3$ \\
 6&$  2.81$&$  0.14$&$  2.76$&$ 98.4$ \\
 7&$ 19.19$&$  0.19$&$  2.76$&$ 14.4$ \\
 8&$  9.60$&$  0.12$&$  2.76$&$ 28.7$ \\
 9&$ 12.01$&$  0.13$&$  2.76$&$ 23.0$ \\
10&$  9.27$&$  0.12$&$  2.76$&$ 29.8$ \\
11&$  9.51$&$  0.17$&$  2.76$&$ 29.0$ \\
\hline
    \end{tabular}
\end{table}

Not only stellar wind, but also radiation and thermal pressure of hot gas can contribute to superbubble expansion. Direct radiation pressure $P_\mathrm{rad}$ can be estimated from the bolometric luminosity $L_\mathrm{bol}$ as \citep{Lopez2014}:

\be {P_\mathrm{rad}/k_B} = \frac{3L_\mathrm{bol}}{4\pi R^2ck_\mathrm{B}} \ee

Following \cite{Barnes2021}, we can estimate $L_\mathrm{bol} \approx 88\mathrm{-}138 L_\mathrm{H\alpha}$ (see figure 7 there). 
To calculate the warm gas thermal pressure, we use the equation:

\be {P_\mathrm{therm}} = (n_\mathrm{e} + n_\mathrm{H} + n_\mathrm{He})T \approx 2n_\mathrm{e}T_\mathrm {e}{k_\mathrm{B}}, \ee

Where $n_\mathrm{e}$ is electron density in a superbubble. Coefficient 2 assumes that all He is singly ionized. We cannot estimate neither $T_\mathrm{e}$ nor $n_\mathrm{e}$ directly from our data. For rough estimate we assume that $n_\mathrm{e} = n_0$ (given in Table~\ref{tab:bubble_params}), and this should provide us a lower limit for $P_\mathrm{therm}$. Similarly, we adopt the value of $T_\mathrm e = 10\,000\ \mathrm{K}$ for warm gas in the superbubbles as a lower limit, given that at low metallicity $T_\mathrm{e}$ can be 1.5--2 times higher \revone{\citep[see, e.g.,][]{PM2017}}.


Finally, we calculate the relative contribution of these terms to the total internal pressure of a superbubble. According to the \cite{Weaver1977} model, evolution of the internal pressure of a bubble can be described as 
\be P_\mathrm{b} = 1.83\times10^5 k_\mathrm{B} L_{38}^{0.4}n_0^{0.6}t_6^{-0.8}. \ee
The results are summarized in Table~\ref{tab:pressure_energy}. As follows from it, the relative fraction of the warm gas thermal pressure in the overall superbubble pressure onto the ISM is mostly within $\sim 25-45$ per cent. More realistic models accounting for the radiation loss or turbulent nature of the ISM suggest the lower value of final $P_\mathrm{b}$ \cite[e.g.][]{Lancaster2021}, and thus resulting contribution of the thermal pressure is probably higher. This implies that the warm ionized gas can significantly contribute to the expansion of the superbubbles. At the same time, radiation pressure is unimportant at this stage, in consistency with findings by \cite{Lopez2014, Barnes2021, McLeod2021}.

Thus, we can conclude that pre-SN feedback in the form of stellar winds and thermal pressure by warm ionized gas (in particular, that leaked through the ISM inhomogeneities from bright \HII regions) together can provide sufficient support for the expansion of most superbubbles in Sextans~A. This solves the problem of the young age of most of the superbubbles \revone{(see Sec.~\ref{sec:supernovae})}. At the same time, as follows from BPT diagram in Fig.~\ref{fig:bpt_img}, the observed line fluxes ratios in superbubbles are better explained if one assumes the contribution of the shocks (probably -- from SNe) to their excitation.

\subsection{Evidence of the impact of wind from the very young low-metallicity star onto ISM?}
\label{sec:disc_star}

The complex~D contains the youngest OB stars discovered by \citet{Garcia2019} (see also \citealt{Lorenzo2022}). One of them -- S004 (see names in their paper) -- has attracted our particular attention. \cite{Garcia2019} classified this star is of O6 Vz spectral type; it has mass  $\sim 25-40 \mathrm{M}_\odot$ and $M_\mathrm{v} = -5.02$. \cite{Lorenzo2022} reassigned its luminosity class to III. 

Our long-slit spectrum covers the star S004 - its position is $\sim60$~arcsec in Fig.~\ref{fig:line_spectra}. The second plot in this Figure demonstrates the zoom-in map of the \Ha velocity dispersion according to the scanning  FPI observations. The increase of $\sigma(\mathrm{H\alpha})$ in the vicinity of this star and a peak associated with it is noticeable there. 
From the fitting of the \Ha line profile (\#3 in Fig.~\ref{fig:i-sigma}), we obtained $\sigma(\mathrm{H\alpha}) = 21 \kms$ that is $\sim 3-4 \kms$ higher than in the vicinity. 
The second highest ratio of \OIIIHb\, in the galaxy is observed toward this star.
We examine the long-slit spectrum in this location and didn't find any peculiarities -- the distribution of lines fluxes ratios agrees well with the other \HII regions, as well as the position of this star on BPT diagrams (shown by crosses in Fig.~\ref{fig:bpt_ls}. The elevation of \SIIHa\ line ratio around the star (Fig.~\ref{fig:line_spectra}) is probably due to the DIG emission dominating there. Thus, we can exclude the shock wave as the source of the elevated velocity dispersion around this star. A possible explanation of it is that we observe a local wind-driven bubble around this very young and metal-poor star, or maybe stellar ejecta. It is interestingly to note that in Sextans~A we expect to see much weaker stellar winds than at higher metallicities \citep[e.g][]{Vink2001}, and the fact that we can directly see its impact onto the \Ha velocity dispersion implies that the wind is not necessarily weak there. This object definitely deserves further observational and theoretical studies.

\subsection{Stellar feedback-driven outflows}
\label{sec:disc_outflow}

As was demonstrated in Section~\ref{sec:velocities}, two regions in Sextans~A (S3 and S7) are especially remarkable by significant blue-shifted velocities of the ionized gas, which are very different from those for the atomic hydrogen. We suggested that we observe an outflowing ionized gas in these regions.

As follows from the analysis of the emission lines flux ratios (Section~\ref{sec:ion_state}), the region S3 is probably density-bounded. \revone{Indeed, the highest ratio of the \OIIIHb\, ratio in the galaxy is observed towards the centre of this superbubble, while \SIIHa\ is not elevated there as one would expect in the case of shock excitation. Such a local maxima of [O~\textsc{iii}]/[S~\textsc{ii}] are typical for optically-thin nebulae \citep[e.g.][]{Pellegrini2012}}. This is an additional indication of the possible stellar feedback-driven outflow in this region.

Roughly assuming a spherical symmetry for the outflow in the region S3, we can estimate the mass-loss rate ($\dot{M}_\mathrm{out}$) and compare this with the SFR in this region of the galaxy (brightest \HII region in the complex~B). \be
\dot{M}_\mathrm{out} = 4\pi R^2\mu m_H<n_\mathrm{p}>V_\mathrm{exp},
\ee
where $<n_\mathrm{p}>$ is the average number density of the ionized hydrogen atoms, which can be significantly different from the density of the neutral hydrogen. We can estimate $<n_\mathrm{p}>$ from the \Ha surface
brightness $I(\mathrm{H\alpha})$ in the same way as in \cite{Egorov2021}, but assuming the total optical path $L = 0.4R$ (since we are measuring $I(\mathrm{H\alpha})$ in the centre of the superbubble S3). 
\be 
<n_\mathrm{p}> \simeq 2.33\times10^8\sqrt{I(\mathrm{H\alpha})/R},
\ee
where $I(\mathrm{H\alpha})$ is in erg~cm$^{-2}$~s$^{-1}$~arcsec$^{-2}$ and $R$ is in pc.
Thus, we obtain $\dot{M}_\mathrm{out} = 5.6\times10^{-3} \mathrm{M}_\odot$~yr$^{-1}$ for the region S3. 
The SFR can be estimated based on the number of ionizing photons produced by the massive stars in S3 and in the adjacent brightest \HII region:  $Q^0 = 1.25\times10^{50}$~s$^{-1}$. According to \citep{Calzetti2010} (eq. 5) for \citep{Kroupa2001} initial mass function, we obtained for S3:
\be \mathrm{SFR}(\mathrm{M}_\odot~\mathrm{yr}^{-1}) = 7.41\times 10^{-54} Q^0_\mathrm{stars} = 9.2\times10^{-4} \mathrm{M}_\odot\ \mathrm{yr}^{-1}.\ee
Using these values, we obtain the mass-loading factor $\eta = \dot{M}_\mathrm{out}/\mathrm{SFR} \simeq 6$ that is slightly higher than the values measured by \cite{McQuinn2019} for several dwarf galaxies with a stellar mass similar to Sextans~A, and consistent with the values predicted in simulations by \cite{Christensen2016}.

The ionized gas kinematics of another region (S7, surrounding the complex~C) \revone{deviates even more} from that of atomic hydrogen, however, the S/N of our \SII\, and \OIII\, images is insufficient to trace the variations of the gas ionization state in this region. 
As was estimated in Sec.~\ref{sec_an_energy}, the superbubble S7 requires about $100-140$ supernovae for its creation. Meanwhile, the number of currently observed OB stars in the complex~C is rather small. If we suggest that the expansion of the complex is driven by SNe explosions, then star formation there should be much more active during the previous 4.6~Myr (kinematical age of the region) than now. \cite{Camacho2016} found that the youngest stars in this complex are older than in the brightest complex~B, while the age of the complex is significantly lower. This means that the active star formation in the complex~C was probably initiated only recently and the duration of a burst of star formation in this complex was short. While we clearly see the blue-shifted velocities of the ionized gas in the region S7 (while these of the atomic gas are mostly red-shifted), the current star formation activity is not sufficient to further support its expansion/outflow.

\section{Summary}
\label{sec:summary}

We present the results of observations of the nearby very low metallicity galaxy Sextans~A performed with a scanning Fabry-Perot interferometer and a long-slit spectrograph at the 6-m BTA telescope (SAO RAS) accompanied by the narrow-band images in \SII\, and \OIII\, lines obtained at the 2.5-m telescope of SAI MSU. 
Our analysis was focused mainly on the kinematics of the ionized gas, whose 2D distribution was never explored before, nor its connection with the morphology and kinematics of the atomic gas, ionization state of ISM, and young massive stars. Our main findings are summarized below:

\begin{itemize}
    \item Kinematics of the ionized gas of Sextans~A cannot be explained by pure circular rotation and is regulated by the stellar feedback from three star-forming complexes in the galaxy. To less extent, this is also true for the neutral gas. \revone{In several regions, kinematics of the ionized and neutral gas differ significantly implying that they are not co-spatial there.}
    \item We identified 11 regions with shell-like morphology of the ionized gas, 10 of them exhibit kinematical evidence of their expansion and thus appear to be ionized gas superbubbles. The remaining regions represent a large structure encircling the brightest star-forming complex in the galaxy by two ionized gas filaments. We found that these filaments have significant line-of-sight velocities in opposite directions and that in contrast with findings by \cite{Hunter1997}, their velocities differ from that in bright \HII regions.
     \item Most of the 10 identified ionized superbubbles are relatively young (1--3~Myr) and have expansion velocities $20-30 \kms$. We estimated the total energy required for their blowing out. Most of the superbubbles of a smaller size can be produced by a few supernovae. For three of them (S1, S2 and S9 in Fig.~\ref{fig:spectr_example}) we found signs of shock excitation in their interior that imply they are supernovae remnants. However, the pre-SN feedback only from the identified young massive stars (in form of winds and thermal pressure of the warm ionized gas) can produce sufficient energy to drive the expansion of most of the superbubbles and thus its contribution to the overall energy balance is significant, despite the very low metallicity environment.
    \item We identified signs of the ionized gas outflow in two regions associated with the brightest star-forming complexes in Sextans~A. Both of them show significant differences between ionized and atomic gas kinematics (ionized gas has blue-shifted velocities). In one of these regions (S3, associated with the star-forming complex B, see Fig.~\ref{fig:spectr_example}) we also identified high \OIIIHb\, lines fluxes ratio pointing to the density-bounded morphology of the region. The estimated mass-loading factor of the outflow in this case is consistent with the simulations by \cite{Christensen2016} for the galaxies of a similar mass. 
    The outflow in another region (S7 -- entire complex C) is probably driven by supernovae and pre-SN feedback from the previous generation of massive stars -- the kinematical age of this region is high ($\sim 5.5$~Myr) and the identified young massive stars produce not sufficient mechanical energy input to support the expansion of this structure.
    \item We found the imprints of feedback (probably -- stellar wind) impacting the surrounding ISM towards one of the youngest O stars identified by \cite{Garcia2019} in the star-forming complex~D. We observe an elevated \Ha velocity dispersion there (by $3-4 \kms$), but do not see any evidence of shock excitation in the emission lines fluxes ratio. While pre-SN feedback is usually considered to be less important in shaping the ISM in a very low metallicity environment, its effects can be still visible in the nearby galaxies. 
\end{itemize}

\section*{Acknowledgements}
We thank Igor Karachentsev and Evgenii  Vasiliev for their useful comments, and Roman Uklein, who performed observations in Feb 2020. We thank the anonymous referee for careful reading of our paper and for providing useful comments and suggestions. 

This study   is based on the data obtained at the   unique scientific facility   the Big Telescope Alt-azimuthal  SAO RAS and  was supported  under  the   Ministry of Science and Higher Education of the Russian Federation grant  075-15-2022-262 (13.MNPMU.21.0003). 
OE acknowledge funding from the Deutsche Forschungsgemeinschaft (DFG, German Research Foundation) in the form of an Emmy Noether Research Group (grant number KR4598/2-1, PI Kreckel). This research made use of Astropy (\url{http://www.astropy.org}) a community-developed core Python package for Astronomy \citep{astropy:2013, astropy:2018} and Astroalign \citep{astroalign}. We acknowledge the usage of the HyperLeda database (\url{http://leda.univ-lyon1.fr}).

\section*{Data Availability}

The data underlying this article will be shared on reasonable request to the corresponding author. The reduced FPI data are available in SIGMA-FPI data base\footnote{\url{http://sigma.sai.msu.ru}} (Egorov et al., in preparation).



\bibliographystyle{mnras}
\bibliography{SexA} 



\appendix
\section{\revone{Quality control of the wavelength solution for the FPI data}}

\revone{In Section~\ref{sec:velocities} we  revealed significant differences in line-of-sight \Ha velocities of the north-western and south-eastern parts of the galaxy (see Fig.~\ref{fig:vel_map}). Because of the importance of this finding for the presented analysis, it is worth checking that there are no systematic differences between the calibration of the individual data cubes that might affect the final mosaic.}

\begin{figure*}
    \centering
    \includegraphics[width=\linewidth]{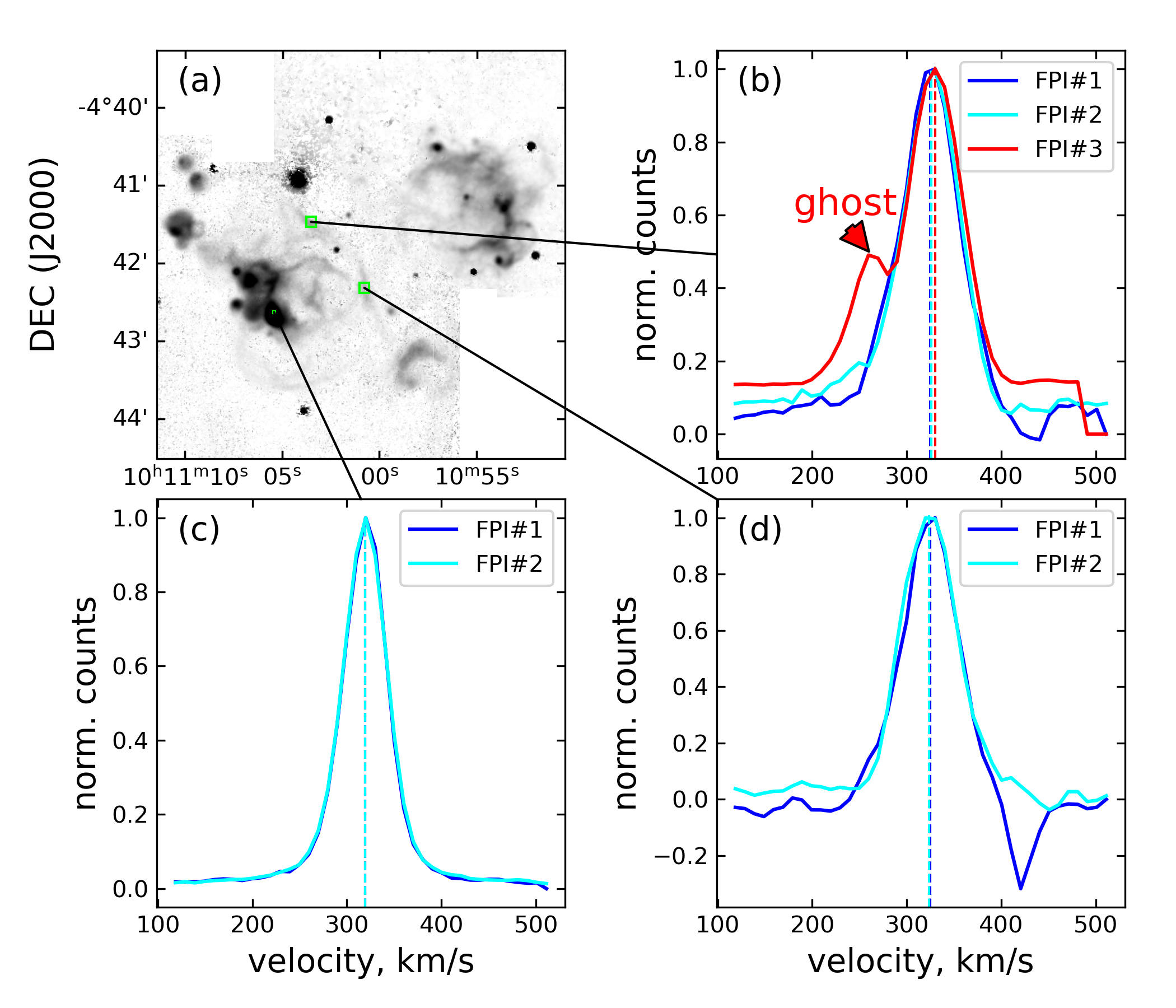}
\caption{\revone{\Ha line profile extracted from two or three different FPI data cubes (b,c,d) within the lime square aperture overlaid on the \Ha flux map (from FPI data; panel a).}}
    \label{fig:check_cubes}
\end{figure*}

\revone{Typical precision of the line-of-sight measurements with the used FPI (IFP751) is $\sim2-3 \kms$ if $S/N \gtrsim 10$
\citep{Moiseev2015}. However, systematic offsets between different data cubes up to $7-8\kms$ are possible. In order to check that our final data cube is not affected by this, we consider the line profiles in the individual data cubes in several regions where the FPI fields overlap. 
There is only one relatively bright region in the overlapping area that can be used to check the consistency of the wavelength solution between all three data cubes. In Fig.~\ref{fig:check_cubes} we demonstrate the \Ha line profiles of the brightness filament in the overlapping area. In this region, the derived velocities are in agreement (within $\sim 5 \kms$). We also considered two other regions where only two data cubes overlap -- the measured velocities there are in even better agreement. This validates the consistency of the wavelength calibration -- possible observational effects are much lower than the differences found in Section~\ref{sec:velocities}. We note also that slight flux variations between different FPI fields are present due to the light scattering at the edges of each FPI field, though this does not affect our analysis of the ionized gas kinematics.}

\bsp	
\label{lastpage}
\end{document}